\documentclass[aps,prc,amsmath,showkeys,superscriptaddress,nofootinbib]{revtex4-1} 
\usepackage{graphicx,epsfig,epstopdf} 

\begin{document}

\title{Many-channel microscopic theory of resonance states \\ and scattering processes
in $^{9}$Be and $^{9}$B}

\author{Yu. A. Lashko}
\email{ylashko@gmail.com}
\affiliation{Bogolyubov Institute for Theoretical Physics,\\
 Kyiv, 03143, Ukraine
}
\affiliation{
National Institute for Nuclear Physics,Padova Division, Padova, Italy}

\author{V. S. Vasilevsky}
\email{vsvasilevsky@gmail.com}
\affiliation{Bogolyubov Institute for Theoretical Physics,\\
 Kyiv, 03143, Ukraine
}
\author{V. I. Zhaba}
\email{viktorzh@meta.ua}
\affiliation{Bogolyubov Institute for Theoretical Physics,\\
 Kyiv, 03143, Ukraine
}

\date{\today }


\begin{abstract}
We present a many-channel microscopic model that extends the three-cluster model previously formulated in \cite{2009NPA...V37}. This extended model incorporates multiple three-cluster configurations, which are subsequently reduced to a comprehensive set of binary channels. These channels dictate the dynamics of various nuclear processes and the resonance structure of a compound nucleus across a broad energy spectrum. The application of this model focuses on investigating the nature of high-energy resonance states in $^{9}$Be and $^{9}$B, as well as the astrophysical $S$-factors for the reactions $^{7}$Li$(d,n)\alpha\alpha$ and $^{7}$Be$(d,p)\alpha\alpha$, particularly pertinent to the cosmological lithium problem.

Parameterization of resonance states is performed across a wide range of total angular momenta and includes states of both positive and negative parity. Dominant decay channels are identified for each resonance state. Detailed analysis of astrophysical $S$ factors resulting from deuteron interactions with $^{7}$Li and $^{7}$Be is conducted within an energy range from zero to 2 MeV. Four exit channels in $^{9}$Be ($^{8}$Be($0^{+}$)$+n$, $^{8}$Be($2^{+}$)$+n$, $^{5}$He($3/2^{-}$)$+\alpha$, $^{5}$He($1/2^{-}$)$+\alpha$) and four in $^{9}$B ($^{8}$Be($0^{+}$)$+p$, $^{8}$Be($2^{+}$)$+p$, $^{5}$Li($3/2^{-}$)$+\alpha$, $^{5}$Li($1/2^{-}$)$+\alpha$) are considered. 

A clear hierarchy of reactions is established for the energy range 0$\leq E<$1.0 MeV. Notably, reactions $^{7}$Li$+d=^{8}$Be($0^{+}$)$+n$ and $^{7}$Be$+d=^{8}$Be($0^{+}$)$+p$ substantially dominate over all other reactions within this energy range. The model satisfactory describes the experimental astrophysical $S$ factors for these reactions.
\end{abstract}


\maketitle

\section{Introduction}

Our main aim is to study the nature of high-energy resonance states in $^{9}$Be and $^{9}$B near $^{7}$Li$+d$ and $^{7}$Be$+d$ decay thresholds and to reveal the influence of these states on the astrophysical $S$-factors of
the reactions $^{7}$Li$\left( d,n\right)
\alpha\alpha$ and $^{7}$Be$\left( d,p\right)  \alpha\alpha$\ 
related to the
cosmological lithium problem. The problem is an overestimation of primordial $^{7}$Li abundance in the standard Big-Bang Nucleosynthesis (BBN) model compared to the observations (see, e.g., \cite{2008JCAP...11..012C}).

In Table \ref{Tab:PrimodAbund},  extracted from Ref. \cite{2016JPhCS.665a2001C} the abundance of certain light nuclei is presented.
\begin{table}[tbph] \centering
\caption{Primordial abundance of light nuclei} 
\begin{ruledtabular} 
\begin{tabular}
[c]{ccccc}
Source & \cite{2010JPhCS.202a2001C} & \cite{2013arXiv1307.6955C,
2014JCAP...10..050C} & \multicolumn{2}{c}{Observations}\\\hline
$Y_{p}$ & 0.2476$\pm$0.0004 & 0.2463$\pm$0.0003 & 0.2534$\pm$0.0083 &
\cite{2012JCAP...04..004A}\\
$D/$H ($\times$10$^{-5}$) & 2.68$\pm$0.15 & 2.67$\pm$0.09 & 3.02$\pm$0.23 &
\cite{2012MNRAS.426.1427O}\\
$^{3}$He$/$H ($\times$10$^{-5}$) & 1.05$\pm$0.04 & 1.05$\pm$0.03 & 1.1$\pm
F$0.2 & \cite{2002Natur.415...54B}\\
$^{7}$Li$/$H ($\times$10$^{-5}$) & 5.14$\pm$0.50 & 4.89$_{-0.39}^{+0.41}$ &
1.58$\pm$0.31 & \cite{2010AxA...522A..26S}\\
\end{tabular}
\end{ruledtabular} 
\label{Tab:PrimodAbund}%
\end{table}%
Notably, the predicted primordial  abundance of $^{7}$Li is about three times higher than its observed value, 
in contrast to the well-reproduced abundance of other light nuclei. 

$^{7}$Li nuclei are primarily formed through the electron capture decay of $^{7}$Be following the conclusion of the primordial BBN. 
So, the reactions which leads to the destruction of $^{7}$Be, such as the $^{7}$Be($d,p$)$^{8}$Be, the $^{7}$Be($n,\alpha$) or $^{7}$Be($n,p$) reactions considered in \cite{2016PhRvL.117o2701B}, could play a part in the reduction of the primordial abundance of $^{7}$Li.

There are several recent measurements of the $^{7}$Be$+d$ reaction.
The first estimate of the $^{7}$Be(d,p)2$\alpha$ reaction rate relied on an extrapolation made by \cite{1972ApJ...175..261P} based on experimental data at center-of-mass energies of 0.6 to 1.3 MeV from \cite{1960NucPh..18..492K}. In this experiment, protons corresponding to the $^8$Be $0^+$ ground state and first excited state (3.03 MeV, $2^+$) were detected at 90$^{\circ}$. 
Assuming an isotropic angular distribution, \cite{1972ApJ...175..261P} multiplied the measured differential cross section by $4 \pi$ and by a further factor of 3 to take into account the estimated contribution of the higher energy $^8$Be states, not observed by \cite{1960NucPh..18..492K}. Consequently, a constant $S$-factor of 100 MeV-barn was adopted.

Ref. \cite{2005ApJ...630L.105A} provides the first direct experimental data for the $^{7}$Be(d,p)2$\alpha$ cross-section at BBN energies (for $T = 0.5 - 1$ GK, the Gamow window is $E = 0.11 - 0.56$ MeV).  Center-of-mass energy ranges of 1.00 to 1.23 MeV  and  0.13 to 0.38 MeV were investigated, and the cross section measurement was averaged over these energy ranges.
The $^{7}$Be(d,p)2$\alpha$ cross section was measured up to an excitation energy in $^8$Be of $E_x = 11.5$ MeV in the BBN related energy range, which allowed observation of the $^7$Be+d reaction via a very broad $4^+$ state ($\Gamma \simeq 3.5$ MeV) situated at an excitation energy of 11.4 MeV in $^8$Be \cite{2004NuPhA.745..155T}. The data \cite{2005ApJ...630L.105A} reveal that the higher energy states  contribute about 35\% to the total $S$-factor instead of the 300\% estimated by \cite{1972ApJ...175..261P},leading to a rate for $^7$Be(d,p)$^8$Be reaction that is a factor of 10 smaller at BBN energies than earlier measurements suggested \cite{2004NuPhA.745..155T}.

In Ref. \cite{2020JPhCS1643a2049I} the $^{7}$Be($d,p$)$^{8}$Be
reaction  has been measured  with a $^{7}$Be target. The
reaction has been reconstructed from the measurement of the outgoing proton alone without measuring the two alpha particles, while maintaining a sufficient energy resolution. Analyzing the cross section of the $^{7}$Be($d,p$)$^{8}$Be reaction, they made a preliminary conclusion that there is no indication of a large resonance at the BBN energy region to destroy $^{7}$Be sufficiently to resolve the cosmological $^{7}$Li problem.

In \cite{2019PhRvL.122r2701R} cross-section of the $d+^7$Be reactions in the
Gamow-window of BBN has been measured. The authors reported that the  cross-section at all energies is dominated by the $\mathrm{^7Be(d,\alpha)^5Li(p)^4He}$ reaction, in contrast to the assumption of Angulo \cite{2005ApJ...630L.105A}, which analyzed cross sections assuming $(d,p)$ kinematics. The cross sections exhibit a resonance behaviour with
dominant peak at $E_{c.m.}$ = 1.17-MeV resonance energy observed in \cite{1960NucPh..18..492K, PhysRevC.84.014308} and a new resonance at E$_{c.m.}$ = 0.36(5)~MeV. The authors attribute such a resonance enhancement of the cross-section to a 5/2$^+$ compound-nuclear state in
$^{9}$B \cite{2008JCAP...11..012C, 2011PhRvD..83f3006C}, an isospin-mirror to the 16.671
(5/2$^+$) state in $^9$Be. Candidates for such a state in $^9$B were
reported at 16.80(10) MeV \cite{PhysRevC.84.014308} and at 16.71 MeV \cite{2012IJMPE..2150004C}.

In \cite{2019PhRvL.122r2701R} BBN mass fractions of ($^7$Li/H)$_P$ = $ 4.66 - 4.69 \cdot 10^{-10}$ is predicted without $d+^7$Be reaction, comparing to ($^7$Li/H)$_P$ =$ 4.24 - 4.61 \cdot 10^{-10}$ in the case of including the latter reaction. The authors concluded that the reduction of the $^7$Li abundance due to the resonance at 0.36(5) MeV could be from 1.4\% to 8.1\%.

So, it follows from the above-mentioned discussion that close to the $d+^7$Be threshold there may be resonances in $^9$B compound nucleus that can affect the astrophysical $S$-factors of the reactions $^{7}$Be$\left( d,p\right)  \alpha\alpha$. The energy and widths of these resonances, their nature and their contribution to the destruction of $^{7}$Be is still a challenging issue. 

New experimental data on high-energy resonance states in $^{9}$Be near $^8$Li$+p$ threshold  were obtained in Ref. \cite{2018PhRvC..98f4601L} and are summarized in Table \ref{Tab:9BeExpNewResons}.%
\begin{table}[tbph] \centering
\caption{Resonance properties (energies in MeV, widths in keV) obtained from $R$-matrix fit. Energies are given with respect to the
$^8$Li +$p$ threshold (16.888 MeV).}%
\begin{ruledtabular} 
\begin{tabular}
[c]{ccccccccc}
\multicolumn{6}{c}{New \cite{2018PhRvC..98f4601L}} &
\multicolumn{3}{c}{Previuos \cite{2004NuPhA.745..155T}}\\\hline
$E$ & $J^{\pi}$ & $\Gamma_{p}$ & $\Gamma_{\alpha}$ & $\Gamma_{d}$ &
$\Gamma_{d^{\prime}}$ & $E$ & $J^{\pi}$ & $\Gamma$\\\hline
0.42$\pm$0.007 & 5/2$^{-}$ & 40$\pm$10 & 20$\pm$3 & 150$\pm$7 &  & 0.41$\pm
$0.007 & (5/2)$^{-}$ & 200\\
0.61$\pm$0.007 & 7/2$^{+}$ & 1.0$\pm$0.2 & 39$\pm$4 & 7$\pm$3 &  & 0.605$\pm
$0.007 & (7/2)$^{+}$ & 47\\
1.10$\pm$0.03 & 3/2$^{+}$ & 10$\pm$7 &  & 30$\pm$10 & 10$\pm$5 & 1.13$\pm
$0.05 &  & \\
1.65$\pm$0.04 & 7/2$^{-}$ & 185$\pm$10 & 185$\pm$12 & 95$\pm$7 & 30$\pm$5 &
1.69$\pm$0.04 &  & \\
1.80$\pm$0.04 & 5/2$^{-}$ & 20$\pm$5 & 14 $\pm$2 & 25$\pm$5 & 20$\pm$5 &
1.76$\pm$0.05 & (5/2$^{-}$) & 300$\pm$100\\
\end{tabular}
\end{ruledtabular} 
\label{Tab:9BeExpNewResons}%
\end{table}%
The study involved three reactions a radioactive $^8$Li beam on a proton target,  analyzed using the R-matrix theory. Refining  positions of the resonances listed in Table \ref{Tab:9BeExpNewResons} and determining their partial widths, the authors found a four-body $\alpha + t + p + n$ resonance near  $18.55\pm 0.04$ MeV in $^9$Be from $^8$Li$(p,d)^7$Li data. They observed a strong $^7$Li$+d$ clustering in resonances  at 0.42 MeV and 1.65 MeV above the $^8$Li+p threshold. So, theoretical interpretation of the nature of these resonance states within cluster models accounting several cluster configurations is in high demand.

\begin{figure}
\begin{center}
\includegraphics[width=\textwidth]{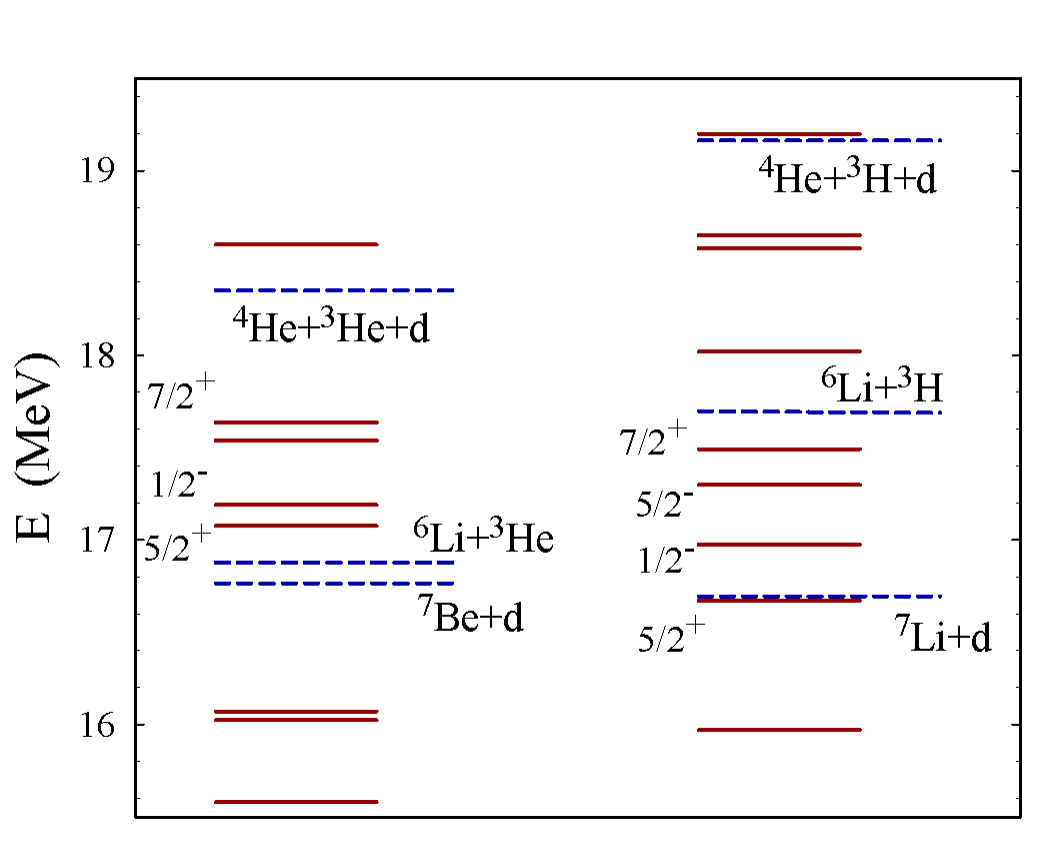}
\caption{Experimental spectrum \cite{2004NuPhA.745..155T} of resonance states  in $^{9}$Be and $^{9}$B
which lie close to the $^{7}$Li+$d$ and $^{7}$Be+$d$ thresholds.}%
\label{Fig:Spectr9Be9BExp}
\end{center}
\end{figure}

We are going to study nature of resonance states around the decay thresholds
$^{7}$Li$+d$, $^{6}$Li$+^{3}$H \ in $^{9}$Be and $^{7}$Be$+d$, $^{6}$Li$+^{3}
$He in $^{9}$B, as the most narrow resonance states have been discovered at
these energy range (see Fig. \ref{Fig:Spectr9Be9BExp} below and Tables 9.2 and 9.13 in Ref. \cite{2004NuPhA.745..155T}). 
Such narrow resonances might substantially change
cross sections of the reactions and affect the $^7$Li abundance \cite{2012IJMPE..2150004C, 2011PhRvC..84d2801O}. Besides, in this region quantum numbers of some experimentally observed states have not been well established yet and require theoretical confirmation. Also theoretical estimation of the partial decay widths of the above-mentioned resonances in $^{9}$Be and $^{9}$B  would be helpful for understanding the cluster structure of these states. 

To study resonances in the $^9$Be, $^9$B nuclei, as well as the astrophysical $S$-factors of the reactions $^{7}$Li$(d,n)\alpha\alpha$ and $^{7}$Be$(d,p)\alpha\alpha$, we expand our models formulated in Ref. \cite{2009NuPhA.824...37V} \ (see also Refs. \cite{2009PAN....72.1450N,
2012PAN.75.818V, 2017NuPhA.958...78L}). We will use the
abbreviation AMGOB denoting the algebraic model of three-cluster systems
utilizing Gaussian and oscillator bases. Within this new model, we  consider reactions using a many-configurational approach. 

The experimental energy difference between the threshold energies for $\alpha+\alpha+n$
and $\quad\alpha+^{3}$H$+d$ channels is 17.5893 MeV for $^{9}$Be, and for
$\alpha+\alpha+p$ and $\alpha+^{3}$He$+d$ channels is 18.353 MeV for $^{9}$B.
While many theoretical investigations focused on the bound state of $^{9}$Be and the low-lying resonance states in $^{9}$Be and $^{9}$B, assuming the latter three-cluster configuration (see, for instance, Refs. \cite{2014PAN..77.555N, 2017PhRvC..96c4322V} and references therein), it is crucial to note that in the region near $^{7}$Be$+d$ and $^{7}$Li$+d$ thresholds, the $\alpha+^{3}$He$+d$ and $\alpha+^{3}$H$+d$ cluster configurations of $^{9}$Be and $^{9}$B become important and should be taken into account.

We take into account the following three-cluster configurations:
\begin{equation}
\alpha+\alpha+n,\quad\alpha+^{3}\text{H}+d \label{eq:I001}%
\end{equation}
in $^{9}$Be and
\begin{equation}
\alpha+\alpha+p,\quad\alpha+^{3}\text{He}+d \label{eq:I002}%
\end{equation}
in $^{9}$B. This allows us to suggest a realistic description of energy spectrum of $^{9}$Be and $^{9}$B in a wide range of energy, where many decay channels of these nuclei are open. With such three-cluster configurations we are able to involve many binary channels, such as
\[
^{8}\text{Be}+n,^{5}\text{He}+\alpha,^{7}\text{Li}+d\text{ and }^{6}%
\text{Li}+t
\]
in $^{9}$Be and%
\[
^{8}\text{Be}+p\text{, }^{5}\text{Li}+\alpha\text{,}^{7}\text{Be}+d\text{ and
}^{6}\text{Li}+^{3}\text{He}%
\]
in $^{9}$B. Moreover, these three-cluster configurations enable us to provide
a more realistic description of  $^{7,8}$Be, $^{5-7}$Li and $,^{5}$He  nuclei considering them as two-cluster systems:
$\alpha+\alpha$, $\alpha+n$, $\alpha+p$, $\alpha+^{3}$H, $\alpha+^{3}$He,
$\alpha+d$, respectively. All these nuclei exhibit a distinct cluster structure, which must be taken into account.

The paper is structured as follows: In Section \ref{model}, we present the main aspects of the microscopic three-cluster model, including the dynamical equations and the asymptotic behavior of the wave functions. Subsection \ref{opt} focuses on the choice of optimal model parameters. In Subsection \ref{res}, we discuss the primary factors responsible for creating or destroying resonance states in a many-channel continuum, while their partial widths are listed in Subsection \ref{part}. The effects of the Coulomb interaction are explored in Subsection \ref{coulomb}, and we provide a comparative analysis of resonance state spectra in $^{9}$Be and $^{9}$B in Subsection \ref{general}. Subsection \ref{sfactor} delves into astrophysical $S$ factors resulting from deuteron interactions with $^{7}$Li and $^{7}$Be, comparing them with experimental data. Finally, our conclusions are summarized in Section \ref{concl}.

\section{Model formulation}
\label{model}
The microscopic model of three-cluster systems, utilizing  Gaussian and oscillator bases, is based on the algebraic version of the resonating group method.

According to the resonating group method, the wave function for the three-cluster system composed of $s$-shell clusters 
($A_{\alpha}\leq4$) can be written as
\begin{eqnarray}
& \Psi^{J}   =\sum_{c}\sum_{\alpha}\label{eq:001}\\
&  \times\widehat{\mathcal{A}}\left\{  \left[  \Phi_{1}^{\left(  c\right)
}\left(  A_{1}^{\left(  c\right)  },S_{1}^{\left(  c\right)  }\right)
\Phi_{2}\left(  A_{2}^{\left(  c\right)  },S_{2}^{\left(  c\right)  }\right)
\Phi_{3}^{\left(  c\right)  }\left(  A_{3}^{\left(  c\right)  },S_{3}^{\left(
c\right)  }\right)  \right]  _{S^{\left(  c\right)  }}\left[  f_{\alpha
}^{\left(  c\right)  }\left(  \mathbf{x}_{\alpha}^{\left(  c\right)
},\mathbf{y}_{\alpha}^{\left(  c\right)  }\right)  \right]  _{L^{\left(
c\right)  }}\right\}  _{J}\nonumber
\end{eqnarray}
where sum over $c$ runs over all three-cluster configurations ($c$=$^{4}$He+$^{4}$He+$n$, $^{4}$He+$^{3}$H+$d$ for $^{9}$Be and $c$=$^{4}$He+$^{4}$He+$p$, $^{4}$He+$^{3}$He+$d$ for $^{9}$B), $\Phi_{\alpha}^{\left(c\right)}\left(  A_{\alpha
},S_{\alpha}\right)  $ is a shell-model wave function for the internal motion
of nucleons within cluster $\alpha$ ($\alpha=1,2,3$), $\ f_{\alpha
}^{\left(c\right)}\left( \mathbf{x}^{\left(c\right)}_{\alpha},\mathbf{y}^{\left(c\right)}_{\alpha}\right)  $ is a Faddeev
component and $S^{\left(c\right)}_{\alpha}$ denotes the spin of the clusters. We use
the $LS$-coupling scheme, i.e.\ total spin $S^{\left(c\right)}$ (a vector sum of the cluster spins $S^{\left(c\right)}_{\alpha}$) couples to the total angular momentum $L^{\left(c\right)}$ and to total momentum $J $.

Vector $\mathbf{x}_{\alpha}^{\left(c\right)}$ is the Jacobi vector, proportional to the distance between $\beta$
and $\gamma$ clusters, while $\mathbf{y}_{\alpha}^{\left(c\right)}$ is a Jacobi vector
connecting the $\alpha$ cluster to the center of mass of the $\beta$\ and
$\gamma$ clusters within cluster configuration $c$:
\label{eq:002}%
\begin{eqnarray}
\mathbf{x}_{\alpha}^{\left(c\right)}  &  =&\sqrt{\frac{A_{\beta}A_{\gamma}}{A_{\beta}+A_{\gamma
}}}\left(  \frac{1}{A_{\beta}}\sum_{j\in A_{\beta}}\mathbf{r}_{j}-\frac
{1}{A_{\gamma}}\sum_{k\in A_{\gamma}}\mathbf{r}_{k}\right)  \qquad
\label{eq:002a}\\
\mathbf{y}_{\alpha}^{\left(c\right)}  &  =& \sqrt{\frac{A_{\alpha}\left(  A_{\beta}+A_{\gamma
}\right)  }{A_{\alpha}+A_{\beta}+A_{\gamma}}}\left(  \frac{1}{A_{\alpha}}%
\sum_{i\in A_{\alpha}}\mathbf{r}_{i}-\frac{1}{A_{\beta}+A_{\gamma}}\left[
\sum_{j\in A_{\beta}}\mathbf{r}_{j}+\sum_{k\in A_{\gamma}}\mathbf{r}%
_{k}\right]  \right)  \label{eq:002b}%
\end{eqnarray}
Indices $\alpha$, $\beta$\ and $\gamma$\ form a cyclic permutation of 1, 2
and 3. The clusters corresponding to indices $\alpha$, $\beta$, and $\gamma$ are determined by the cluster configuration $c.$ Below, we will sometimes omit index $c$ to avoid making the formulas too cumbersome.

For each Faddeev component within fixed cluster partition $c$ we use bi-spherical harmonics
\begin{equation}
f_{\alpha}\left(\mathbf{x}_{\alpha},\mathbf{y}_{\alpha}\right)  \Rightarrow
f_{\alpha}^{\left(  L\right)  }\left( \mathbf{x}_{\alpha},\mathbf{y}_{\alpha
}\right)  =\sum_{\lambda_{\alpha},l_{\alpha}}f_{\alpha}^{\left(
\lambda_{\alpha},l_{\alpha};L\right)  }\left(  x_{\alpha},y_{\alpha}\right)
\left\{  Y_{\lambda_{\alpha}}\left(  \widehat{\mathbf{x}}_{\alpha}\right)
Y_{l_{\alpha}}\left(  \widehat{\mathbf{y}}_{\alpha}\right)  \right\}  _{LM}
\label{eq:003}%
\end{equation}
which lead to four quantum numbers $\lambda_{\alpha},l_{\alpha},LM$.  In
$s$-shell clusters  the angular momentum $L$ is completely
determined by the inter-cluster motion. The parity of the three-cluster states
is determined by the partial angular momenta $\lambda_{\alpha},\,l_{\alpha}$ associated with Jacobi vectors $\mathbf{x}_{\alpha},\mathbf{y}_{\alpha},$ correspondingly: $\pi=\left(-\right)^{\lambda_{\alpha}+l_{\alpha}}$.

It is well know that Faddeev components are very suitable for implementing the
necessary boundary conditions both for binary and for three-cluster decay channels of the nucleus under consideration (see \cite{kn:faddeev+merk93} for more details).

\subsection{Asymptotic behavior for Faddeev amplitudes}
\label{asymp}
In this subsection, we will omit index $c$, assuming that we are considering a fixed cluster configuration.
To solve set of equations for Faddeev amplitudes $f_{\alpha}^{\left(  \lambda_{\alpha},l_{\alpha};L\right)  }\left(
x_{\alpha},y_{\alpha}\right)$, we need to determine their asymptotic behavior.  There are four asymptotic regimes in
asymptotic regions which were denoted as $\Omega_{\alpha}$ ($\alpha$=1, 2, 3)
and $\Omega_{0}$ by Faddeev and Merkuriev \cite{kn:faddeev+merk93}. The region $\Omega_{0}$
corresponds to the case when all three particles are well separated, and all
distances are large: $x_{\alpha}\gg1,\,y_{\alpha}\gg1$. In contrast, $\Omega_{\alpha}$  correspond to the case when two particles with indices $\beta$ and $\gamma$ are close to each other and form
a bound state(s) characterized by energy $\mathcal{E}^{\sigma}_{\alpha}$ and function(s) $\psi_{{\mathcal{E}^{\sigma}_{\alpha}},\lambda_{\alpha}}^{\alpha}\left(  
x_{\alpha}\right)$, while the third particle with index $\alpha$ is far away. It
is obvious that in the region $\Omega_{\alpha}$, the\ coordinate $y_{\alpha}$
is much larger than the coordinate $x_{\alpha}$.  

In the asymptotic region $\Omega_{\alpha}$,
component $f_{\alpha}^{\left(  \lambda_{\alpha},l_{\alpha};L\right)  }\left(
x_{\alpha},y_{\alpha}\right)  $ of the total wave function has an asymptotic form%
\[
f_{\alpha}^{\left(  \lambda_{\alpha},l_{\alpha};L\right)  }\left(  x_{\alpha
},y_{\alpha}\right)  \approx \psi_{{\mathcal{E}^{\sigma}_{\alpha}},\lambda_{\alpha}}^{\alpha}\left(  
x_{\alpha}\right)\left[  \delta_{i_{0},i_{\alpha}}\Psi_{l_{\alpha}}^{\left(-\right)}\left(
q_{\alpha}y_{\alpha}\right)  -S_{i_{0},i_{\alpha}}\Psi_{l_{\alpha}}^{\left(
+\right)  }\left(  q_{\alpha}y_{\alpha}\right)  \right]
\]
where $\Psi_{l_{\alpha}}^{\left(-\right)}$ is the incoming wave and 
$\Psi_{l_{\alpha}}^{\left(+\right) }$ is the outgoing wave connected with regular
$F_{l_{\alpha}}\left(  \eta_{\alpha},q_{\alpha}y_{\alpha}\right)  $ and
irregular $G_{l_{\alpha}}\left(  \eta_{\alpha},q_{\alpha}y_{\alpha}\right)  $
Coulomb functions \cite{1983RvMP...55..155B} by the relation%
\[
\Psi_{l_{\alpha}}^{\left(  \pm\right)  }=F_{l_{\alpha}}\left(  \eta_{\alpha
},q_{\alpha}y_{\alpha}\right)  \pm iG_{l_{\alpha}}\left(  \eta_{\alpha
},q_{\alpha}y_{\alpha}\right).
\]
Wave number
$q_{\alpha}$ and Sommerfeld parameter $\eta_{\alpha}$ are defined as%
\begin{align*}
q_{\alpha}  &  =\sqrt{\frac{2m\left(  E-\mathcal{E}^{\sigma}_{\alpha}\right)  }{\hbar^{2}}},\\
\eta_{\alpha}  &  =\frac{Z_{\alpha}\left(  Z_{\beta}+Z_{\gamma}\right)  e^{2}%
}{\sqrt{2\left(  E-\mathcal{E}_{\sigma\alpha}\right)  }}\sqrt{\frac{m}{\hbar^{2}}%
\frac{A_{\alpha}\left(  A_{\beta}+A_{\gamma}\right)  }{A}},
\end{align*}
where $Z_{\alpha}$, $Z_{\beta}$ and $Z_{\gamma}$ are the charges of individual
clusters, $E$ is the total energy of three-cluster system ($E>\mathcal{E}^{\sigma}_{\alpha}$) and $m$ is the mass of a nucleon.

The elements of the scattering matrix $S_{i_{0},i_{\alpha}}$ determine the transition
from the incoming channel $i_{0}$ to the outgoing channel $i_{\alpha}$. The multiple
index$\ i_{\alpha}$ contains complete information about the present channel: it
includes the orbital momentum $l_{\alpha}$ of a "projectile", its energy
$E-\mathcal{E}^{\sigma}_{\alpha}$, as well as the orbital momentum $\lambda_{\alpha}$  of a "target" (two-cluster subsystem): $i_{\alpha}=\left\{\mathcal{E}^{\sigma}_{\alpha},l_{\alpha},\lambda_{\alpha}\right...\}$

\subsection{Dynamical equations}
The wave function of a three-cluster system can be represented through the set of binary channels:%
\begin{equation}
\Psi^{J}=\sum_{c}\sum_{\alpha}\widehat{\mathcal{A}}\left\{\left[ \psi_{j_{\alpha}^{(c) },\lambda_{\alpha}^{(c) }}^{(c,\alpha)}\left(  
\mathbf{x}_{\alpha}^{(c) }\right) \Phi_{\alpha}^{(c) }\left(  A_{\alpha}^{\left(
c;\right)  },S_{\alpha}^{\left(  c\right)  }\right)   \phi_{l_{\alpha}^{(c)}}^{\left(c,\alpha\right)}\left(  \mathbf{y}_{\alpha}^{\left(  c\right)  }\right)
\right]  \right\}  _{J} \label{eq:004}%
\end{equation}
where%
\begin{align}
&  \psi_{j_{\alpha}^{(c) },\lambda_{\alpha}^{(c) }}^{(c,\alpha)}\left(  
\mathbf{x}_{\alpha}^{\left(  c\right)  }\right) \label{eq:005}\\
&  =\widehat{\mathcal{A}}\left\{\left[ \left[  \Phi_{\beta}^{\left(
c\right)  }\left(  A_{\beta}^{\left(  c\right)  },S_{\beta}^{\left(  c\right)
}\right)  \Phi_{\gamma}\left(  A_{\gamma}^{\left(  c\right)  },S_{\gamma
}^{\left(  c\right)  }\right)  \right]_{S_{\beta+\gamma}^{\left(  c\right)
}}\chi^{(c,\alpha)}_{\lambda_{\alpha}^{\left(  c\right)  }}\left(  \mathbf{x}_{\alpha
}^{\left(  c\right)  }\right)  \right] _{j_{\alpha}^{\left(  c\right)
}}\right\} \nonumber
\end{align}
is a wave function of two-cluster subsystem composed of clusters $\beta$ and $\gamma$. The binary subsystem is characterized by 
angular momentum $j_{\alpha}^{\left(  c\right)}$ which is a vector sum of orbital momentum $\lambda_{\alpha}^{\left(  c\right)}$ and spin $S_{\beta+\gamma}.$ Function
$\phi_{l_{\alpha}^{(c)}}^{\left(c,\alpha\right)}\left(  \mathbf{y}_{\alpha}^{\left(  c\right)  }\right)$ describes relative motion of cluster $\alpha$ and two-cluster subsystem $(\beta+\gamma)$  with 
orbital momentum of the relative motion $l_{\alpha}^{\left(  c\right)}.$

The microscopic Hamiltonian for a three-cluster configuration $c$ is given by
a sum of three single-cluster Hamiltonians  $\hat{H}_\alpha^{(1)}$ describing the internal structure of each cluster, and a term responsible for the inter-cluster dynamics: 
\begin{equation}
\hat{H}=\hat{T}+\hat{V}=\sum_{\alpha=1}^3\hat{H}_\alpha^{(1)}+\hat{T}_r+\sum_{\alpha}\hat{V}_\alpha    
\end{equation}
The latter consists of the
kinetic energy operator $\hat{T}_r$ for relative motion of clusters and  potential energy of interaction $\hat{V}_\alpha$  between clusters:
\begin{equation*}
\hat{H}_\alpha^{(1)}=\sum_{i\in A_\alpha}\hat{T}(i)+\sum_{i<j\in A_\alpha}\hat{V}(ij),\,
\hat{V}_\alpha=\sum_{i\in A_\beta}\sum_{j\in A_\gamma}\hat{V}(ij),\,
\hat{T}_r=-\frac{\hbar^2}{2m}\left(\Delta_{\mathbf{x}_{\alpha}}+\Delta_{\mathbf{y}_{\alpha}}\right)
\end{equation*}
The Hamiltonian $\hat{H}$ can be also expressed in an equivalent form 
\begin{equation}
\hat{H}=\hat{H}_\alpha^{(1)}+\hat{H}_\alpha^{(2)}+\hat{T}_\alpha+\sum_{\beta\neq\alpha}\hat{V}_\beta
\label{Hamiltonian}
\end{equation}
as the sum of the two-cluster Hamiltonian $\hat{H}_\alpha^{(2)},$ the Hamiltonian of the third cluster 
$\hat{H}_\alpha^{(1)}$, and the Hamiltonian representing the interaction of the third cluster with the binary subsystem:
\begin{equation*}
\hat{H}_\alpha^{(2)}=\sum_{i\in A_\beta+A_\gamma}\hat{T}(i)+\sum_{i<j\in A_\beta+A_\gamma}\hat{V}(ij),\,
\hat{T}_\alpha=-\frac{\hbar^2}{2m}\Delta_{\mathbf{y}_{\alpha}}
\end{equation*}

As in \cite{2017NuPhA.958...78L}, we use a finite number of square-integrable Gaussian functions to expand the two-cluster wave function $\psi_{j_{\alpha}^{(c) },\lambda_{\alpha}^{(c) }}^{(c,\alpha)}\left(  
\mathbf{x}_{\alpha}^{\left(  c\right)  }\right)$
\begin{align}
& \psi_{j_{\alpha}^{(c) },\lambda_{\alpha}^{(c) }}^{(c,\alpha)}\left(  
\mathbf{x}_{\alpha}^{\left(  c\right)  }\right)=
\sum_{\nu=1}^{N_{\alpha}^{G}}D_{j_{\alpha}^{(c)},\lambda_{\alpha}^{(c) },\nu }^{(c,\alpha)}\,\chi_{j_{\alpha}^{(c)},\lambda_{\alpha}^{(c) },\nu }^{(c,\alpha)}\\
&
=\sum_{\nu=1}^{N_{\alpha}^{G}}D_{j_{\alpha}^{(c)},\lambda_{\alpha}^{(c) },\nu }^{(c,\alpha)}\widehat{\mathcal{A}}\left\{\left[ \left[  \Phi_{\beta}^{\left(
c\right)  }\left(  A_{\beta}^{\left(  c\right)  },S_{\beta}^{\left(  c\right)
}\right)  \Phi_{\gamma}\left(  A_{\gamma}^{\left(  c\right)  },S_{\gamma
}^{\left(  c\right)  }\right)  \right]_{S_{\beta+\gamma}^{\left(  c\right)
}}G_{\lambda_{\alpha}^{(c)}}\left(  x_{\alpha
}^{( c)},b_{\nu}\right) Y_{\lambda_{\alpha}}\left(  \widehat{\mathbf{x}}_{\alpha}\right)\right] _{j_{\alpha}^{\left(  c\right)
}}\right\},\nonumber
\label{eq:117}%
\end{align}
where
\begin{eqnarray}
G_{\lambda}\left( x,b_{\nu}\right) =
\sqrt{\frac{2}{b_{\nu}^{3}\Gamma\left(  \lambda_{\alpha}+3/2\right)  }}\rho
^{\lambda_{\alpha}}\exp\left\{  -\frac{1}{2}\rho^{2}\right\},  \label{eq:118} 
\rho=\frac{x}{b_{\nu}}   \nonumber %
\end{eqnarray}
is a Gaussian function. Parameters $b_\nu$ are chosen so to minimize the ground state energies of the two-body subsystems.

Expansion coefficients $D_{j_{\alpha}^{(c)},\lambda_{\alpha}^{(c) },\nu }^{(c,\alpha,\sigma)}$ are solutions of the two-cluster equation:
\begin{equation}
\sum_{\nu=1}^{N_{\alpha}^{G}}\langle \nu, \alpha, c| \hat{H}_\alpha^{(2)}-\mathcal{E}_{\sigma}^{\alpha,c}|{\tilde \nu}, \alpha, c\rangle D_{j_{\alpha}^{(c)},\lambda_{\alpha}^{(c) },\nu }^{(c,\alpha,\sigma)}=0  
\end{equation}
These coefficients are a discrete analogue of the two-cluster wave function $\psi_{{\mathcal{E}^{\sigma}_{\alpha}},j_{\alpha},\lambda_{\alpha}}^{\alpha}\left(  
x_{\alpha}\right)$ and
$\mathcal{E}_{\sigma}^{\alpha,c}$ is energy of bound or pseudo-bound state of the two-cluster system.

The total wave function of the three-cluster system is expanded in a Gaussian basis and a complete set of oscillator states so that 
\begin{equation}
\Psi^{J}=\sum_{c}\sum_{\alpha}\sum_{\lambda_{\alpha},l_{\alpha}}\sum_{\nu_{\alpha},n_{\alpha}} C_{\nu_{\alpha}\lambda_{\alpha};n_{\alpha}l_{\alpha}}^{(c,\alpha)}
\widehat{\mathcal{A}}\left\{\left[ \chi_{j_{\alpha}^{(c) },\lambda_{\alpha}^{(c) },\nu_{\alpha}}^{(c,\alpha)}\left(  
\mathbf{x}_{\alpha}^{(c) }\right) \Phi_{\alpha}^{(c) }\left(  A_{\alpha}^{\left(
c;\right)  },S_{\alpha}^{\left(  c\right)  }\right)   \varphi_{n_{\alpha}l_{\alpha}^{(c)}}^{\left(c,\alpha\right)}\left(  \mathbf{y}_{\alpha}^{\left(  c\right) },b\right)
\right]  \right\}  _{J} \label{eq:007}%
\end{equation}
where%
\begin{eqnarray*}
\varphi_{n_{\alpha}l_{\alpha}^{(c)}}^{\left(c,\alpha\right)}\left(  \mathbf{y}_{\alpha}^{\left(  c\right) },b\right)=\varphi_{n_{\alpha}l_{\alpha}^{(c)}}^{\left(c,\alpha\right)}\left(y_{\alpha}^{\left(  c\right) },b\right)Y_{\lambda_{\alpha}}\left(  \widehat{\mathbf{y}}_{\alpha}\right); \end{eqnarray*}
\begin{eqnarray}
\varphi_{n_{\alpha},l_{\alpha}}\left( y_{\alpha},b\right)= \left(
-1\right)  ^{n_{\alpha}}\mathcal{N}_{n_{\alpha}l_{\alpha}}%
{\tilde \rho}^{l_{\alpha}}e^{-\frac{1}{2}{\tilde \rho}^{2}}L_{n_{\alpha}}^{l_{\alpha}
+1/2}\left({\tilde \rho}^{2}\right)  ,\quad\label{eq:121}\\
{\tilde \rho}  = \frac{y_{\alpha}}{b}, \quad\mathcal{N}_{n_{\alpha}l_{\alpha}}%
=\sqrt{\frac{2\Gamma\left(  n_{\alpha}+1\right)  }{b^{3}~\Gamma\left(  n_{\alpha
}+l_{\alpha}+3/2\right)  }}\nonumber
\end{eqnarray}
is an oscillator function and $b$ is the oscillator length.

To obtain the dynamic equations for the total wave function $C_{\nu_{\alpha}\lambda_{\alpha};n_{\alpha}l_{\alpha}}^{(c,\alpha)}$ in the discrete
representation, we substitute the expansion (\ref{eq:007}) in the Schrödinger equation containing Hamiltonian (\ref{Hamiltonian}):
\begin{equation}
\sum_{c}\sum_{\alpha}\sum_{\lambda_{\alpha},l_{\alpha}}\sum_{\nu_{\alpha},n_{\alpha}}
\langle {\tilde \nu_{\alpha}}, {\tilde\lambda_\alpha};{\tilde n_\alpha},{\tilde l_\alpha};{\tilde\alpha}, {\tilde c}| \hat{H}-E|\nu_{\alpha},\lambda_{\alpha};n_{\alpha},l_{\alpha};\alpha, c\rangle C_{\nu_{\alpha}\lambda_{\alpha};n_{\alpha}l_{\alpha}}^{(c,\alpha)}=0
\label{matrix}
\end{equation}
This equation should be supplemented by the appropriate boundary conditions. The latter have been discussed in the previous section. Actually, we reduce multi-configuration three-cluster equation to a set of coupled two-body equations. This means that we treat a three-cluster scattering as a scattering of the third cluster on the two-cluster subsystem being in the state characterized by energy $\mathcal{E}_{\sigma}^{\alpha,c}.$  So, we make unitary transformation of the three-cluster Hamiltonian matrix entering Eq. (\ref{matrix}) to the matrix in the representation of pairs of interacting clusters:
\begin{align}
&\langle {\tilde \sigma}, {\tilde\lambda_\alpha};{\tilde n_\alpha},{\tilde l_\alpha};{\tilde\alpha}, {\tilde c}| \hat{H}-E|\sigma, \lambda_{\alpha};n_{\alpha}, l_{\alpha};\alpha, c\rangle=\nonumber\\
&\sum_{\nu,{\tilde \nu}}D_{{\tilde j_{\alpha}^{(c)}},{\tilde\lambda_{\alpha}^{(c) }},{\tilde \nu} }^{({\tilde c},{\tilde\alpha},{\tilde\sigma})}\langle {\tilde \nu_{\alpha}}, {\tilde\lambda_\alpha};{\tilde n_\alpha},{\tilde l_\alpha};{\tilde\alpha}, {\tilde c}| \hat{H}|\nu_{\alpha},\lambda_{\alpha};n_{\alpha},l_{\alpha};\alpha, c\rangle D_{j_{\alpha}^{(c)},\lambda_{\alpha}^{(c) },\nu }^{(c,\alpha,\sigma)}
\label{matrix_new}
\end{align}
As a result, we obtain a set of equation for coefficients $C^{(i_{\alpha})}_{{ n_\alpha},l_\alpha},$ where index $i_{\alpha}$ denotes all the quantum numbers labelling different binary decay channels of the three-cluster system studied (see subsection \ref{asymp}).

\begin{equation}
\sum_{i_{\alpha}}\sum_{{n_\alpha}=0}^{\infty}\langle {\tilde \sigma}, {\tilde\lambda_\alpha};{\tilde n_\alpha},{\tilde l_\alpha};{\tilde\alpha}, {\tilde c}| \hat{H}-E|\sigma, \lambda_{\alpha};n_{\alpha}, l_{\alpha};\alpha, c\rangle C^{(i_{\alpha})}_{{ n_\alpha,l_\alpha}}=0  
\end{equation}

Matrix elements (\ref{matrix_new}) coupling different channels $i_{\alpha}$ vanish with increasing the number of oscillator quanta $n_\alpha, {\tilde n_\alpha}$ and the following asymptotic relation becomes valid for the expansion coefficients $C^{(i_{\alpha})}_{{ n_\alpha}}:$    


\[
C^{(i_{\alpha})}_{{ n_\alpha, l_\alpha}}  \approx   \delta_{i_{0},i_{\alpha}}\Psi_{l_{\alpha}}^{\left(-\right)}\left(
q_{\alpha}r_{n_{\alpha}}\right)  -S_{i_{0},i_{\alpha}}\Psi_{l_{\alpha}}^{\left(
+\right)  }\left(  q_{\alpha}r_{n_{\alpha}}\right).
\]
Here $r_{n_\alpha}=b\sqrt{4n_\alpha+2l_\alpha+3}$ is a discrete analogue of Jacobi vector ${y}_{\alpha}$ describing the distance between $(\beta+\gamma)$ binary subsystem and the remaining cluster.
\bigskip 

In the $LS$ coupling scheme one can easily formulate selections rules for
possible values of the total orbital momentum L and total spin compound system
$S$ when the total angular momentum $J$ and parity $\pi$\ is fixed. It
necessary to note that for the three-cluster configuration $c=\alpha+\alpha+n$ the total spin $S^{(c)}$  is fully determined by the spin of valence neutron and takes the only value $S^{(c)}$=1/2. For $c=\alpha+t+d$
three-cluster configuration of $^9$Be  or $c=\alpha+^{3}$He$+d$ cluster partition of $^9$B total spin $S^{(c)}$ is a vector sum of a deuteron and a triton or a deuteron and $^3$He, correspondingly. Hence, there are two possible values of the total spin: $S^{(c)}$=1/2 and $S^{(c)}$=3/2. 

Thus for a fixed value
of the total angular momentum $J$ and parity $\pi$ two states with $S^{(c)}$=1/2 with the total orbital momentum
\[
L^{(c)}=J-1/2,L=J+1/2
\]
are to be involved into consideration, while four states
with
\[
L^{(c)}=J-3/2,J-1/2,L^{(c)}=J+1/2,L=J+3/2
\]
are to be taken into account when the total spin $S^{(c)}$=3/2.

The exception to this rule is the state $J^\pi=1/2^-$ which contains the only value of total orbital momentum $L=1$, because it is impossible to construct a negative parity state with $L=0$ for three s-clusters.

At the final stage of constructing matrix elements of Hamiltonian and norm kernel,
we need to transfer from the LS coupling scheme to the JJ scheme. They are
both equvalent, and for bound states give the same results if the
diagonalization procedure is used. For scattering states, it is more
appropriate to employ the JJ coupling scheme.
In the JJ scheme we introduce two new quantum numbers $j_{1}$ and $j_{2}$.
They are related to the vector operators $\mathbf{J}_{1}$ and $\mathbf{J}_{2}%
$. The operator $\mathbf{J}_{2}$ is a vector sum of the operators
\[
\mathbf{J}_{2}=\mathbf{\lambda}+\mathbf{S}_{23}%
\]
where $\mathbf{S}_{23}$ is the spin operator of two-cluster subsystem, and the
quantum number  $j_{2}$ is the  angular momentum of two-cluster subsystem
(target). The operator $\mathbf{J}_{1}$ is determined as a vector sum of the
operators%
\[
\mathbf{J}_{1}=\mathbf{l}+\mathbf{S}_{1}%
\]
and thus the quantum number $j_{1}$ is the angular momentum of the third
cluster (projectile). In the JJ scheme we use the following set of five
quantum numbers distinguishing channels of three-cluster system:%
\begin{equation}
c=\left\{  \lambda,j_{2},l,j_{1};J\right\}
\label{eq:jjchannels}
\end{equation}
There is a simple relation involving 9j symbols that connects cluster functions in the JJ coupling
scheme 
\[\left\vert \nu,n,\lambda,j_{2},l,j_{1};J\right\rangle \] 
and the LS
coupling scheme $\left\vert \nu,n,\lambda,l;L,S,J\right\rangle $%
\begin{eqnarray*}
\left\vert \nu,n,\lambda,j_{2},l,j_{1};J\right\rangle  & =&\sum_{L,S}%
\sqrt{\left(  2S+1\right)  \left(  2L+1\right)  \left(  2j_{2}+1\right)
\left(  2j_{1}+1\right)  }\\
& \times&\left\{
\begin{array}
[c]{ccc}%
S_{23} & \lambda & j_{2}\\
S_{1} & l & j_{1}\\
S & L & J
\end{array}
\right\}  \left\vert \nu,n,\lambda,l;L,S,J\right\rangle.
\end{eqnarray*}
This relation is used to transform from one
coupling scheme to another for matrix elements of all operators employed in
our calculations.

\section{Results and discussion}

To obtain spectrum of resonance states and cross sections of the reactions, we
employ of the Minnesota potential (MP) \cite{kn:Minn_pot1} with the IV version
of the spin-orbital interaction \cite{1970NuPhA.158..529R}.

Our model, as any version of the resonating group method, contains a free
parameter - an oscillator length, which has to be selected in the most
appropriate way. There is another adjustable parameter in our calculations, as well as in others that employ the MP. It is an exchange parameter $u$ determining the intensity of the odd components of the nucleon-nucleon potential. As in Ref.
\cite{2014PAN..77.555N}, where three-cluster resonance states in $^{9}$Be and
$^{9}$B at low-energy region have been studied, we selected the oscillator
length $b$=1.285 fm. This value of $b$ minimizes the energy of three-cluster
threshold $\alpha+\alpha+N$. In Ref. \cite{2014PAN..77.555N} the exchange
parameter $u$ was selected to reproduce the bound state energy of $^{9}$Be
with respect to the three-cluster threshold $\alpha+\alpha+n$. However, in the
present paper, we use a larger part of the total Hilbert space, thus we need to
use another criterion for selecting the parameter $u$.

We employ four Gaussian functions to describe bound and pseudo-bound states of
two-cluster subsystems. As was shown in Refs. \cite{2009NuPhA.824...37V, 
2009PAN....72.1450N, 2012PAN.75.818V, 2017UkrJPh..62..461V, 2017NuPhA.958...78L}, such number of
Gaussian functions provides precise enough energies and wave functions of bound
two-cluster states.  An algorithm for selecting the optimal
parameters of the Gaussian basis was explained in detail in Refs.
\cite{2009NuPhA.824...37V, 2009PAN....72.1450N, 2012PAN.75.818V, 
2017UkrJPh..62..461V, 2017NuPhA.958...78L}. To achieve unitarity of the scattering $S$ matrix with acceptable accuracy, we employ two hundred oscillator functions in our calculations.

\subsection{Optimal parameters}
\label{opt}
\begin{figure}
\begin{center}
\includegraphics[width=\textwidth]{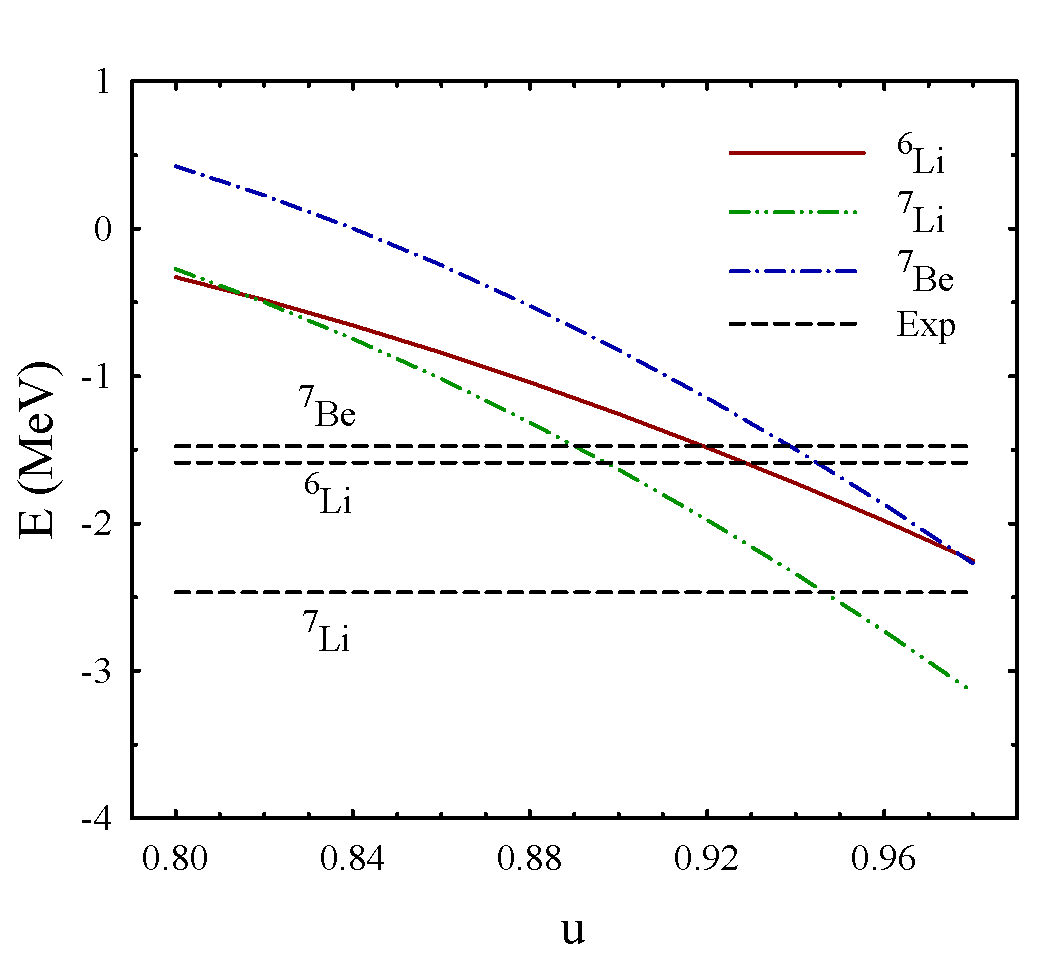}
\caption{Bound state energy of $^{6}$Li,  $^{7}$Li and $^{7}$Be with respect to the $\alpha+d$, $\alpha+t$, $\alpha+^{3}$He thresholds as a function of the parameter $u$. Experimental data are represented by horizontal dashed
lines.}%
\label{Fig:EThreshU}%
\end{center}
\end{figure}

In this paper, we  use a little bit different criteria for selecting optimal
value of the exchange parameter $u$ of the MP. Since our focus is on studying resonance states near the $^{7}$Li+$d$, $^{6}$Li+$^{3}$H thresholds in $^{9}$Be and near the $^{7}$Be+$d$, $^{6}$Li+$^{3}$He thresholds in $^{9}$B, it becomes crucial to offer the most accurate description of the internal structure of nuclei $^{6}$Li, $^{7}$Li, and $^{7}$Be. It means to find
an optimal value of the exchange parameter $u$ to reproduce the ground state
energy of nuclei $^{6}$Li, $^{7}$Li and $^{7}$Be with respect to the
$\alpha+d$, $\alpha+t$ and $\alpha+^{3}$He thresholds, correspondingly. Fig.
\ref{Fig:EThreshU} illustrates the dependence of bound state energies for $^{6}$Li, $^{7}$Li, and $^{7}$Be\ on the parameter $u$. The horizontal dashed lines
indicate the experimental values for these energies. It's evident that there isn't a single value of $u$ that accurately reproduces all three energies.
The optimal value of $u$ is 0.929 for $^{6}$Li, 0.946 for $^{7}$Li, and 0.939 for $^{7}$Be. Thus fitting the ground state energy of
$^{6}$Li, $^{7}$Li and $^{7}$Be suggests that the range for optimal values of
$u$ is 0.929$\leq u\leq$0.946.

\begin{figure}
\begin{center}
\includegraphics[width=\textwidth]{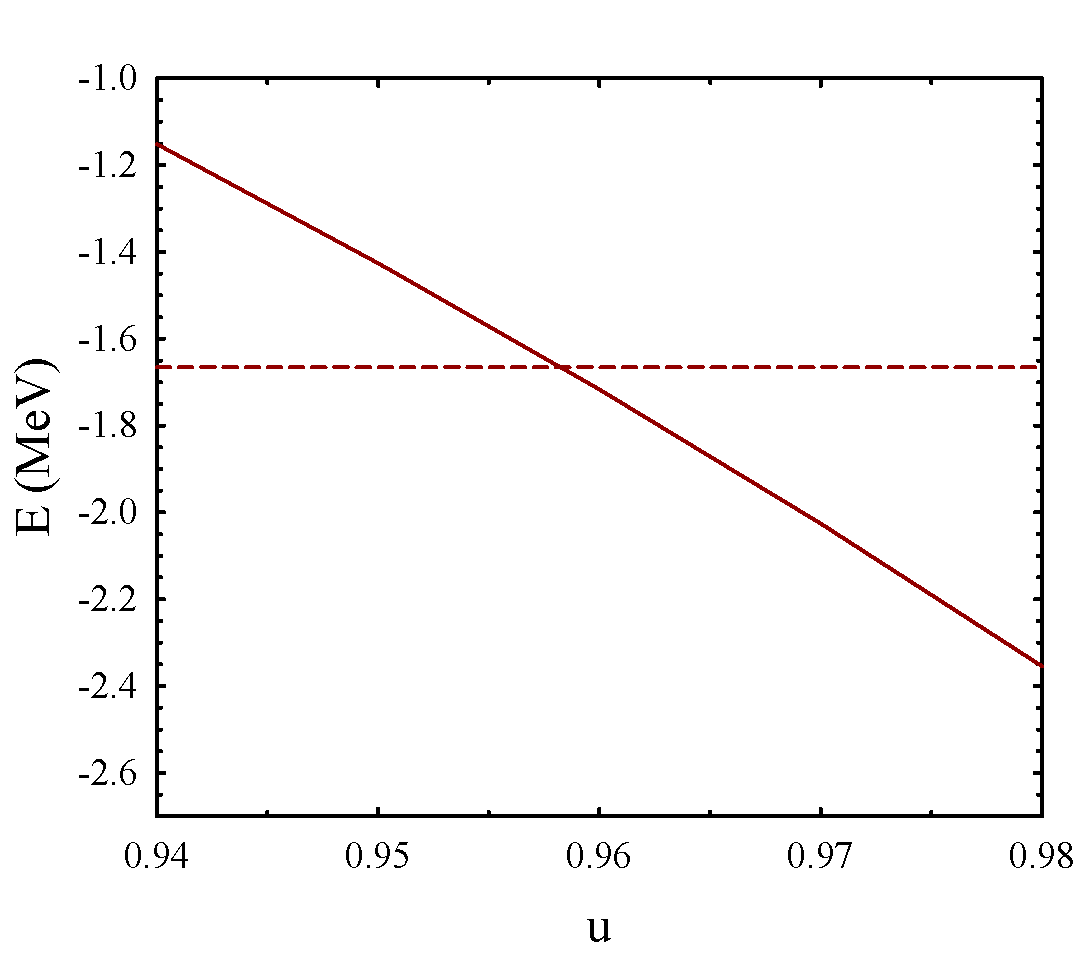}
\caption{Energy of the 3/2$^{-}$ bound state of $^{9}$Be with respect to the $^8$Be$+n$ decay threshold as a function of the
exchange parameter $u$. The dashed horizontal line indicates the experimental
energy.}%
\label{Fig:LowEnerg9BevsU}%
\end{center}
\end{figure}

This range of the parameter $u$ is acceptable
for description of the internal structure of the nuclei $^{6}$Li, $^{7}$Li and
$^{7}$Be. Meanwhile, we need an additional criterion of selecting parameter
$u$ related to intercluster interactions. For this aim we consider the only
bound state of the $^{9}$Be nucleus. Thus now we turn our attention to the
ground state of $^{9}$Be. According to Ref. \cite{2004NuPhA.745..155T}, the
ground state energy is -1.6654\ MeV with respect to the threshold $^{8}%
$Be+$n$. In Fig. \ref{Fig:LowEnerg9BevsU}, we show how energy of the 3/2$^{-}$
ground state depends on the exchange parameter $u$. It is evident that the
optimal value of $u$ for the ground state of $^{9}$Be is close to 0.958. This
value falls outside the range of optimal values for the internal structure of $^{6}%
$Li, $^{7}$Li and $^{7}$Be. Therefore, we have decided on $u$=0.95 as a suitable compromise between describing the internal structure of clusters $^{6}$Li, $^{7}$Li, and $^{7}$Be, and reproducing the bound state energy of the compound system $^{9}$Be. Results of this
compromise is shown in Table \ref{Tab: BoundStatesA6A7A9} where we compare
the calculated and experimental energies of the clusters and $^{9}$Be.
Experimental data are taking from Refs. \cite{2002NuPhA.708....3T,
2004NuPhA.745..155T}. With such a value of $u$, ground state energies
$^{7}$Li and $^{7}$Be are close to the experimental values, while energies of the ground
states of $^{6}$Li and $^{9}$Be are obtained with acceptable precision.%

\begin{table}[htbp] \centering
\caption{Calculated and experimental energies of bound states for $^6$Li, $^7$Li, $^7$Be and $^9$Be. Energies are given in MeV with respect to the $\alpha+d$, $\alpha+t$, $\alpha+^{3}$He, and $^8$Be$+n$ thresholds}%
\begin{ruledtabular} 
\begin{tabular}
[c]{ccc}
Nucleus & Theor. & Exp.\\\hline
$^{6}$Li & -1.289 & -1.475\\
$^{7}$Li & -2.524 & -2.467\\
$^{7}$Be & -1.673 & -1.587\\
$^{9}$Be & -1.542 & -1.665\\
\end{tabular}
\end{ruledtabular} 
\label{Tab: BoundStatesA6A7A9}%
\end{table}%

\subsection{Formation of resonance states}
\label{res}

Now we are going to study how resonance states are formed, or, in other words,
what are the main factors which create or destroy resonance states in
many-channel continuum. 

We start with the 5/2$^{+}$ and 7/2$^{+}$ resonance states in $^{9}$B, as
they were determined experimentally. 

 \begin{table}[htbp] \centering
\caption{Effects of different channels on parameters of the 5/2$^+$  and  7/2$^+$ resonance states in $^9$B. Energies $E$ and widths $\Gamma$ are given in MeV, with energies measured relative to the $^7$Be$+d$ decay threshold. The channels are included in the following order: $^{7}$Be+$d$, $^{6}$Li+$^{3}$He, $^{8}$Be+$p$ and $^{5}$Li+$\alpha$}%
\begin{ruledtabular} 
\begin{tabular}
[c]{cccccccc}
$J^{\pi}$ & Model & $E$ & $\Gamma$ & $E$ & $\Gamma$ & $E$ & $\Gamma$\\\hline
5/2$^{+}$ & 1C & 3.407 & 3.735 & 6.954 & 4.562 &  & \\
& 2C & 2.215 & 2.039 & 5.914 & 4.683 &  & \\
& 3C & 3.466 & 5.104 &  &  &  & \\
& 4C & 3.074  & 6.002 & 3.720 & 6.415 &  & \\
7/2$^{+}$ & 1C & 2.422 & 1.765 &  4.721  & 0.977 & 5.277 & 3.464\\
& 2C & 2.366 & 0.209 &  &  &  & \\
& 3C & 2.098 & 1.322  & 2.423 & 1.547 & 4.055 & 3.205\\
& 4C & 2.083 & 1.320 & 2.282 & 1.460 & 2.439 & 1.578\\
\end{tabular}
\end{ruledtabular} 
\label{Tab:Form52P72P}%
\end{table}%

Table \ref{Tab:Form52P72P} displays the formation of 5/2$^{+}$ and 7/2$^{+}$ resonance states, positioned above the $^{7}$Be+d threshold. We use four approximations: 1C, 2C, 3C, and
4C, which stand for the single-channel ($^{7}$Be+d), two-channel ($^{7}$Be+d and
$^{6}$Li+$^{3}$He), three-channel ($^{7}$Be+$d$, $^{6}$Li+$^{3}$He and $^{8}%
$Be+$p$), and four-channel approximations ($^{7}$Be+$d$, $^{6}$Li+$^{3}$He, $^{8}$Be+$p$ and $^{5}$Li+$\alpha$). One can see that in the $^{7}$Be+d
channel,  the combination of the centrifugal and Coulomb barriers creates a
huge total barrier which allows to accommodate two resonances in the 5/2$^{+}$
state and three resonance states in the 7/2$^{+}$ state. They are all wide
with exception of the second  7/2$^{+}$ state. 

The role of the second channel, $^{6}$Li+$^{3}$He, is particularly intriguing. It reduces noticeable the energy and width
of the first 5/2$^{+}$ resonance state and also the total width of the first
7/2$^{+}$ resonance state. However, this channel dissolves the second and the
third 7/2$^{+}$ resonance states, and slightly increases the width of the
5/2$^{+}$ resonance state. 

Even more intriguing is the role of $^{8}$Be+$p$ and
$^{5}$Li+$^{4}$He channels. The $^{8}$Be+$p$ channel significantly increases
the energy and width of the first 5/2$^{+}$ resonance state and dissolves the
second 5/2$^{+}$ state. Meanwhile, this channel restores two 7/2$^{+}$ resonance states, which were dissolved by the $^{6}$Li+$^{3}$He channel.

Now we consider how different channels affect the 5/2$^{-}$ resonance states
in $^{9}$B. 
\begin{table}[tbph] \centering
\caption{Effects of different channels on parameters of the 5/2$^-$   resonance states in $^9$B. Energies $E$ and widths $\Gamma$ are in MeV, with energies measured relative to the $^7$Be$+d$ decay threshold. The channels are included in the following order: $^{7}$Be+$d$, $^{6}$Li+$^{3}$He, $^{8}$Be+$p$ and $^{5}$Li+$\alpha$}%
\begin{ruledtabular} 
\begin{tabular}
[c]{cccccccccccc}
Model & $E$ & $E$ & $\Gamma$ & $E$ & $\Gamma$ & $E$ & $\Gamma$ & $E$ &
$\Gamma$ & $E$ & $\Gamma$\\\hline
1C & -5.764 & -3.392 & - & -0.860 & - & 2.539 & 0.672 & 3.774 & 6.247 &  & \\
2C & -9.025 & -4.678 & - & -2.824 & - & 1.120 & 0.656 & 2.284 & 0.959 &
3.245 & 2.504\\
3C &  & -4.614 & 0.102 & -1.692 & 2.435 & 1.228 & 0.788 & 2.347 & 1.250 &
3.323 & 2.560\\
4C &  & -4.609 & 0.251 &  &  & 1.205 & 0.938 & 2.372 & 1.351 & 3.306 &
2.561\\
\end{tabular}
\end{ruledtabular} 
\label{Tab:Form12M}%
\end{table}%

In four-channel approximation we detected one resonance state
below the $^{7}$Be+$d$ threshold and three resonances above it. When we use
single-channel (1C) and two-channel (2C) approximations, the energy region
below the $^{7}$Be+$d$ threshold can contain only bound states, as there is no
way for decay of such states. That is why in Table \ref{Tab:Form12M} we
display energies of bound states obtained in 1C and 2C approximations. There
are three bound states created by the $^{7}$Be+$d$  channel and by coupled
$^{7}$Be+$d$  and $^{6}$Li+$^{3}$He channels. By adding one open ($^{8}%
$Be+$p$) channel or two open ($^{8}$Be+$p$ and $^{5}$Li+$^{4}$He) channels, we
dissolve two bound states and  transform one bound state into a narrow
resonance state. The latter state is then the resonance of the Feshbach type,
because it is created due to coupling of closed and open channels.

And now let us consider the shape 5/2$^{-}$ resonance states, i.e. those
resonance states which lie above the $^{7}$Be+$d$ decay threshold. A shape resonance, unlike a Feshbach resonance, remains unbound even when the coupling between different channels is switched off. 
Actually
all shape resonance states shown in Table \ref{Tab:Form12M} lie above both
$^{7}$Be+$d$  and $^{6}$Li+$^{3}$He thresholds. The $^{7}$Be+$d$ channel
alone creates two resonance states, one of them is narrow and another is very
wide. Two channels $^{7}$Be+$d$  and $^{6}$Li+$^{3}$He  together create two
narrow and one fairly wide resonance state. Two additional channels $^{8}$Be+$p$ and $^{5}$Li+$^{4}$He slightly change energies and width of these
resonances. These results indicate that there is a weak coupling of the $^{7}$Be+$d$  and $^{6}$Li+$^{3}$He with  channels $^{8}$Be+$p$ and $^{5}$Li+$^{4}$He in the energy range 0$\leq E\leq$5 MeV above the $^{7}$Be+$d$
decay threshold.

\begin{figure}
\begin{center}
\includegraphics[width=\textwidth]{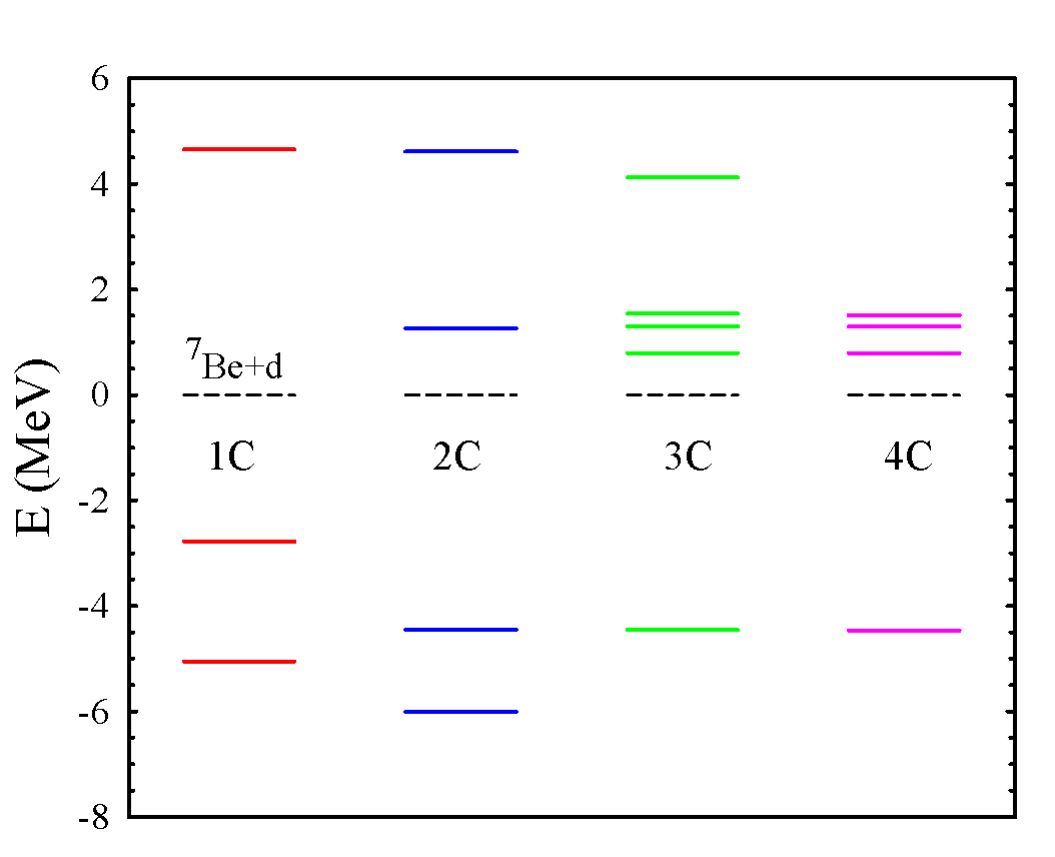}
\caption{Formation of the 1/2$^-$ resonance states in $^{9}$B. The channels are included in the following order: $^{7}$Be+$d$, $^{6}$Li+$^{3}$He, $^{8}$Be+$p$ and $^{5}$Li+$\alpha$}.%
\label{Fig:FormR12M9B}%
\end{center}
\end{figure}

\begin{figure}
\begin{center}
\includegraphics[width=\textwidth]{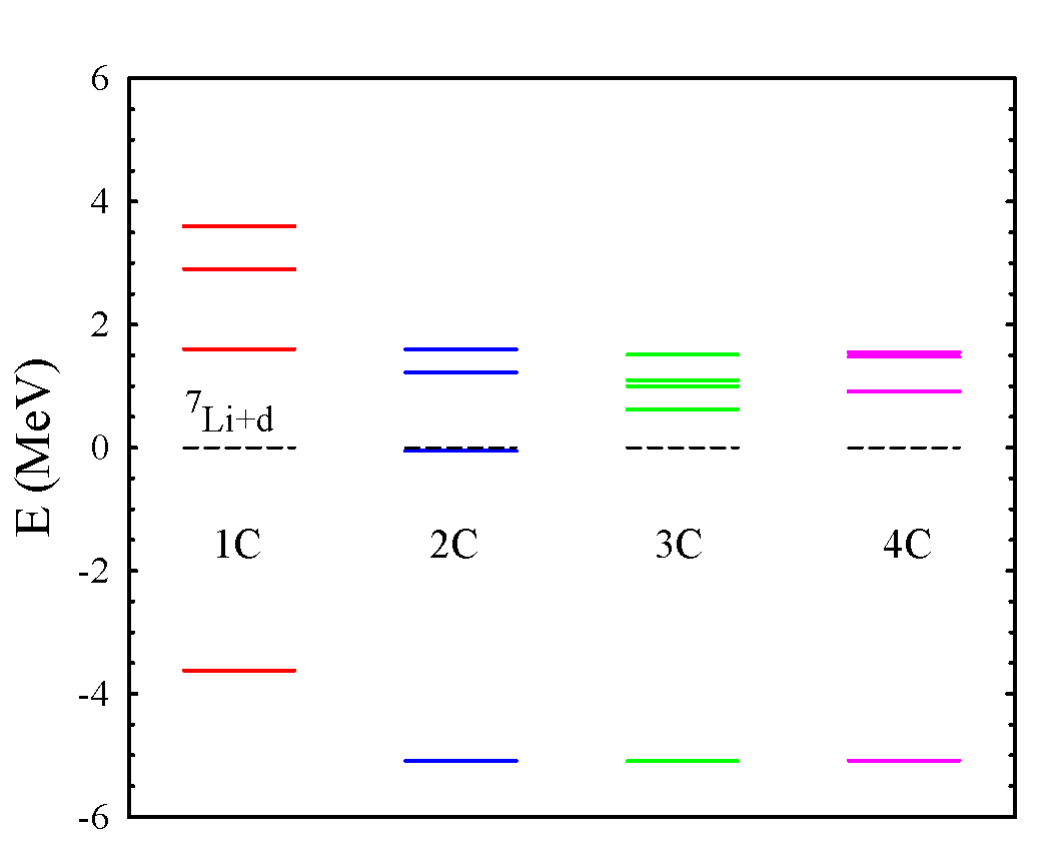}
\caption{Formation of the 3/2$^{-}$ resonance states in $^{9}$Be. The channels are included in the following order: $^{7}$Li+$d$, $^{6}$Li+$^{3}$H, $^{8}$Be+$n$ and $^{5}$He+$\alpha$.}%
\label{Fig:FormR32M9Be}%
\end{center}
\end{figure}

In Figs. \ref{Fig:FormR12M9B} and \ref{Fig:FormR32M9Be} we visualize formation
of the 1/2$^{-}$ resonance states in $^{9}$B and 3/2$^{-}$ resonance states in
$^{9}$Be. In both cases two channels $^{7}$Be+$d$ and $^{6}$Li+$^{3}$He
($^{7}$Li+$d$ and $^{6}$Li+$^{3}$H) create two resonance states and two bound
states. By adding two other channels $^{8}$Be+$p$ and $^{5}$Li+$^{4}$He
($^{8}$Be+$n$ and $^{5}$He+$^{4}$He), one of the bound states is dissolved
and another  is transformed into resonance state with approximately the same
energy. Regarding the spectrum of resonance states above the $^{7}$Be+$d$ ($^{7}$Li+$d$) threshold, it undergoes a dramatic transformation due to the inclusion of the two new channels. Nevertheless, in both cases, the energy of one of the resonance states experiences only a slight alteration.

\subsection{Partial widths}
\label{part}
To understand the nature of the revealed resonance states, we calculate
and analyze the total $\Gamma$ and partial widths $\Gamma_{c}$ ($c$=1,2,
\ldots, $N_{c}$, is the total number of channels for the selected $J^{\pi}$
state). These quantities allow us to reveal  dominant decay channels. Due to the inclusion of a significant number of channels in our model, we will selectively present a limited number of partial widths and quantum numbers associated with the corresponding decay channels. We will also display the ratio $\Gamma
_{c}/\Gamma$, this will immediately indicate the importance of
the  channel $c$.

In Table \ref{Tab:PartWidth9B52M} we present the total and partial widths for
two  5/2$^{-}$ resonance states in $^{9}$B, listing six channels. These channels account for 99.6\% of the total width for the first 5/2$^{-}$ resonance state and 95.1\% for the second resonance state. One of the resonances lies below the $^{7}
$Be+d decay threshold and has a small width ($\Gamma=$ 100 keV), the other
locates above the threshold and is a wide resonance state. One can see that the
$^{4}$He+$^{5}$Li(3/2$^{-}$) channel with the angular momentum of relative
motion $j_{1}=2$ is dominant decay channel of the first resonance state. This
channel exhausts 70\% of the total  decay width, while the second important
channel $p$+$^{8}$Be(2$^{+}$) exhausts only 13\%. Another 5/2$^{-}$ resonance state exhibits a notably different structure. In this resonance state, the dominant channel is $d$+$^{7}$Be(3/2$^{-}$). It accounts for only 41\% of the total width, while the subsequent significant channel $p$+$^{8}$Be(2$^{+}$) contributes 29\%.  The
channel $^{3}$He+$^{6}$Li(1$^{+}$), with the threshold energy close to the
energy of the dominant $d$+$^{7}$Be(3/2$^{-}$) channel, has a relatively small
contribution (7\%) to the total width. 

\begin{table}[tbp] \centering
\caption{Dominant decay channels of the 5/2$^-$ resonance states in $^9$B. Energies, total and partial widths are given in MeV, with energies measured relative to the $^7$Be$+d$ decay threshold. }%
\begin{ruledtabular} 
\begin{tabular}
[c]{cccccccc}
\multicolumn{4}{c}{$J^{\pi}=5/2^{-}$, $E$=-4.165, $\Gamma=$0.100} &
\multicolumn{4}{c}{$J^{\pi}=5/2^{-}$, $E$=1.486, $\Gamma=$1.539}\\\hline
$\Gamma_{c}$ & $\Gamma_{c}/\Gamma,$ \% & Channel & $j_{1}$ & 
$\Gamma_{c}$ & $\Gamma_{c}/\Gamma,$ \% & Channel & $j_{1}$\\\hline
0.070 & 69.91 & $^{4}$He+$^{5}$Li(3/2$^{-}$) & 2  & 0.636 & 41.32 &
$d$+$^{7}$Be(3/2$^{-}$) & 1\\
0.015 & 14.90 & $p$+$^{8}$Be(2$^{+}$) & 3/2  & 0.451 & 29.31 &
$p$+$^{8}$Be(2$^{+}$) & 3/2\\
0.007 & 7.45 & $p$+$^{8}$Be(2$^{+}$) & 1/2  & 0.102 & 6.63 & $^{3}%
$He+$^{6}$Li(1$^{+}$) & 3/2\\
0.005 & 5.35 & $p$+$^{8}$Be(2$^{+}$) & 7/2  & 0.096 & 6.24 & $p$+$^{8}%
$Be(2$^{+}$) & 1/2\\
0.001 & 1.03 & $^{4}$He+$^{5}$Li(1/2$^{-}$) & 2 &  0.091 & 5.91 &
$d$+$^{7}$Be(3/2$^{-}$) & 3\\
0.0009 & 0.96 & $p$+$^{8}$Be(2$^{+}$) & 5/2 &  0.087 & 5.65 & $d$%
+$^{7}$Be(3/2$^{-}$) & 2\\
\end{tabular}
\end{ruledtabular} 
\label{Tab:PartWidth9B52M}
\end{table}

In Table \ref{Tab:PartWidth9Be52M} we demonstrate the total and partial widths
of the 5/2$^{-}$ resonance states in $^{9}$Be. 
Parameters of these resonance states
are close to the parameters of the 5/2$^{-}$ resonance states in $^{9}$B. The
differences between them are mainly due to the Coulomb interaction.%

\begin{table}[tbp] \centering
\caption{Dominant decay channels of the 5/2$^-$ resonance states in $^9$Be. Energies, total and partial widths are given in MeV, with energies measured relative to the $^7$Li$+d$ decay threshold.}%
\begin{ruledtabular} 
\begin{tabular}
[c]{cccccccc}
\multicolumn{4}{c}{$J^{\pi}=5/2^{-}$, $E$=-4.413, $\Gamma=$0.245} &
\multicolumn{4}{c}{$J^{\pi}=5/2^{-}$, $E$=1.479, $\Gamma=$1.263}\\\hline
 $\Gamma_{c}$ & $\Gamma_{c}/\Gamma,$ \% & Channel & $j_{1}$ &
$\Gamma_{c}$ & $\Gamma_{c}/\Gamma,$ \% & Channel & $j_{1}$\\\hline
 0.172 & 69.96 & $^{4}$He+$^{5}$He(3/2$^{-}$) & 2 &  0.640 & 50.65 &
$d$+$^{7}$Li(3/2$^{-}$) & 1\\
 0.061 & 25.00 & $^{4}$He+$^{5}$He(3/2$^{-}$) & 4 &  0.264 & 20.92 &
$n$+$^{8}$Be(2$^{+}$) & 3/2\\
 0.009 & 3.74 & $n$+$^{8}$Be(2$^{+}$) & 3/2 &  0.118 & 9.31 & $d$+$^{7}%
$Li(3/2$^{-}$) & 2\\
 0.002 & 0.87 & $n$+$^{8}$Be(0$^{+}$) & 5/2 &  0.058 & 4.60 & $^{4}%
$He+$^{5}$He(3/2$^{-}$) & 4\\
 0.0006 & 0.25 & $n$+$^{8}$Be(2$^{+}$) & 5/2 &  0.057 & 4.50 & $n$%
+$^{8}$Be(2$^{+}$) & 1/2\\
 0.0003 & 0.11 & $n$+$^{8}$Be(2$^{+}$) & 7/2 &  0.053 & 4.16 & $d$%
+$^{7}$Li(3/2$^{-}$) & 3\\
\end{tabular}
\end{ruledtabular} 
\label{Tab:PartWidth9Be52M}%
\end{table}%

As anticipated, the mirror channels $d$+$^{7}$Li(3/2$^{-}$), $n$+$^{8}$Be(2$^{+}$), and $d$+$^{7}$Be(3/2$^{-}$), $p$+$^{8}$Be(2$^{+}$), exhibit dominance in the resonance states with energies of $E$=1.479 MeV in $^{9}$Be and $E$=1.486 MeV in $^{9}$B. These channels collectively contribute 71.6\% in $^{9}$Be and 70.6\% in $^{9}$B to the total width. Furthermore, two other resonance states with energies of $E$=-4.413 MeV in $^{9}$Be and $E$=-4.165 MeV in $^{9}$B, lying below the $d$+$^{7}$Li and $d$+$^{7}$Be thresholds, respectively, also primarily undergo decay via the mirror channels $^{4}$He+$^{5}$He(3/2$^{-}$) and $^{4}$He+$^{5}$Li(3/2$^{-}$). These channels account for approximately 70\% of the total width in both $^{9}$Be and $^{9}$B. However, it's noteworthy that for these latter resonances, the second most important channels are not mirror channels. In the case of $^{9}$Be, this secondary channel is $^{4}$He+$^{5}$He(3/2$^{-}$) with a 25\% contribution, whereas in $^{9}$B, it's the channel $p$+$^{8}$Be(2$^{+}$) with a 15\% contribution to the total width.

In Table \ref{Tab:PartWidth9Be12M} we show dominant decay channels of \ the
1/2$^{-}$\ and 3/2$^{-}$ resonance state in $^{9}$Be and in Table
\ref{Tab:PartWidth9B12M} we display dominant decay channels of \ the 1/2$^{-}%
$\ and 3/2$^{-}$ resonance state in $^{9}$B. These resonance states lie very
close to the thresholds $d$+$^{7}$Li and $d$+$^{7}$Be, respectively.%

\begin{table}[tbp] \centering
\caption{Dominant decay channels of the 1/2$^-$ and  3/2$^-$ resonance states in $^9$Be. Energies, total and partial widths are given in MeV, with energies measured relative to the $^7$Li$+d$ decay threshold.}%
\begin{ruledtabular} 
\begin{tabular}
[c]{cccccccc}
\multicolumn{4}{c}{$J^{\pi}=1/2^{-}$, $E$=0.448, $\Gamma=$ 0.680} &
\multicolumn{4}{c}{$J^{\pi}=3/2^{-}$, $E$=0.915, $\Gamma=$1.418}\\\hline
 $\Gamma_{c}$ & $\Gamma_{c}/\Gamma$, \% & Channel & $j_{1}$ & 
$\Gamma_{c}$ & $\Gamma_{c}/\Gamma$, \% & Channel & $j_{1}$\\\hline
 0.472 & 69.36 & $n$+$^{8}$Be(0$^{+}$) & 1/2 &  0.525 & 36.99 & $^{3}%
$H+$^{6}$Li(1$^{+}$) & 3/2\\
 0.196 & 28.80 & $d$+$^{7}$Li(3/2$^{-}$) & 2 &  0.391 & 27.55 & $^{3}%
$H+$^{6}$Li(1$^{+}$) & 5/2\\
0.008 & 1.25 & $^{4}$He+$^{5}$He(3/2$^{-}$) & 2 &  0.216 & 15.20 &
$d$+$^{7}$Li(3/2$^{-}$) & 1\\
0.002 & 0.35 & $^{4}$He+$^{5}$He(1/2$^{-}$) & 0 &  0.143 & 10.09 &
$^{3}$H+$^{6}$Li(1$^{+}$) & 1/2\\
 0.001 & 0.22 & $n$+$^{8}$Be(2$^{+}$) & 3/2 &  0.052 & 3.65 & $n$+$^{8}%
$Be(2$^{+}$) & 1/2\\
 0.0001 & 0.01 & $d$+$^{7}$Li(3/2$^{-}$) & 1 &  0.038 & 2.65 & $n$%
+$^{8}$Be(2$^{+}$) & 3/2\\
\end{tabular}
\end{ruledtabular} 
\label{Tab:PartWidth9Be12M}%
\end{table}%

\begin{table}[tbp] \centering
\caption{Dominant decay channels of the 1/2$^-$  and 3/2$^-$ resonance states in $^9$B. Energies, total and partial widths are given in MeV, with energies measured relative to the $^7$Be$+d$ decay threshold. }%
\begin{ruledtabular} 
\begin{tabular}
[c]{cccccccc}
\multicolumn{4}{c}{$J^{\pi}=1/2^{-}$, $E$=0.793, $\Gamma=$1.101} &
\multicolumn{4}{c}{$J^{\pi}=3/2^{-}$, $E$=1.013, $\Gamma=$0.812}\\\hline
$\Gamma_{c}$ & $\Gamma_{c}/\Gamma$, \% & Channel & $j_{1}$  &
$\Gamma_{c}$ & $\Gamma_{c}/\Gamma$, \% & Channel & $j_{1}$\\\hline
 0.654 & 59.45 & $p$+$^{8}$Be(0$^{+}$) & 1/2  & 0.225 & 27.73 & $^{3}%
$He+$^{6}$Li(1$^{+}$) & 3/2\\
 0.436 & 39.61 & $d$+$^{7}$Be(3/2$^{-}$) & 2  & 0.174 & 21.47 & $^{3}%
$He+$^{6}$Li(1$^{+}$) & 5/2\\
 0.004 & 0.34 & $^{4}$He+$^{5}$Li(1/2$^{-}$) & 0  & 0.161 & 19.78 &
$d$+$^{7}$Be(3/2$^{-}$) & 1\\
 0.003 & 0.24 & $^{4}$He+$^{5}$Li(3/2$^{-}$) & 2  & 0.103 & 12.66 &
$^{3}$He+$^{6}$Li(1$^{+}$) & 1/2\\
 0.002 & 0.21 & $^{3}$He+$^{6}$Li(1$^{+}$) & 3/2  & 0.068 & 8.32 &
$p$+$^{8}$Be(2$^{+}$) & 1/2\\
 0.001 & 0.06 & $d$+$^{7}$Be(3/2$^{-}$) & 1 &  0.020 & 2.46 & $p$+$^{8}%
$Be(2$^{+}$) & 3/2\\
\end{tabular}
\end{ruledtabular} 
\label{Tab:PartWidth9B12M}%
\end{table}%

As we see, two channels compete for decay of the 1/2$^{-}$ resonance states in
$^{9}$Be and $^{9}$B. They are $n$+$^{8}$Be(0$^{+}$) and $d$+$^{7}$%
Li(3/2$^{-}$) channels in $^{9}$Be, the first channel exhausts $\approx$70\%
of total width and the second channel exhausts $\approx$29\%. In $^{9}$B two
channels $p$+$^{8}$Be(0$^{+}$) and $d$+$^{7}$Be(3/2$^{-}$) are dominant, the
contribution of the first channel is approximately 60\% and the contribution
of the second channel is $\approx$40\%. Different situation is observed for
the 3/2$^{-}$ resonance states. Four channels give noticeable contribution in
$^{9}$Be, and five channels dominate in $^{9}$B. Their summary contribution is
almost 90\%.

\begin{figure}
\begin{center}
\includegraphics[width=\textwidth]{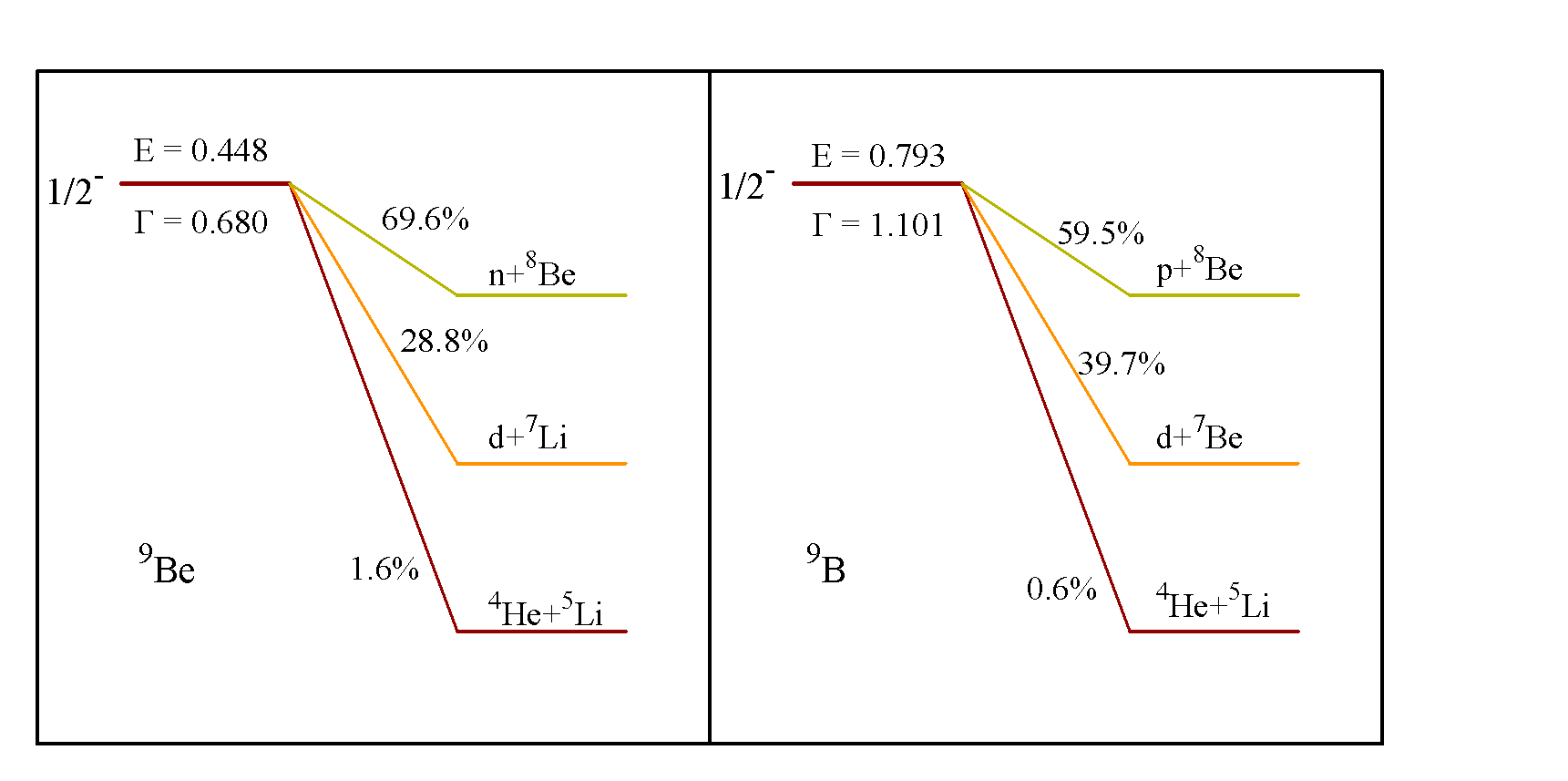}
\caption{Dominant decay channels of the 1/2$^{-}$ resonance states in $^{9}$Be
and $^{9}$B. Energies and widths are given in MeV, with energies measured relative to the $^7$Li$+d$ and $^7$Be$+d$ decay thresholds.}%
\label{Fig:DominDecayChR12M}%
\end{center}
\end{figure}

\begin{figure}
\begin{center}
\includegraphics[width=\textwidth]{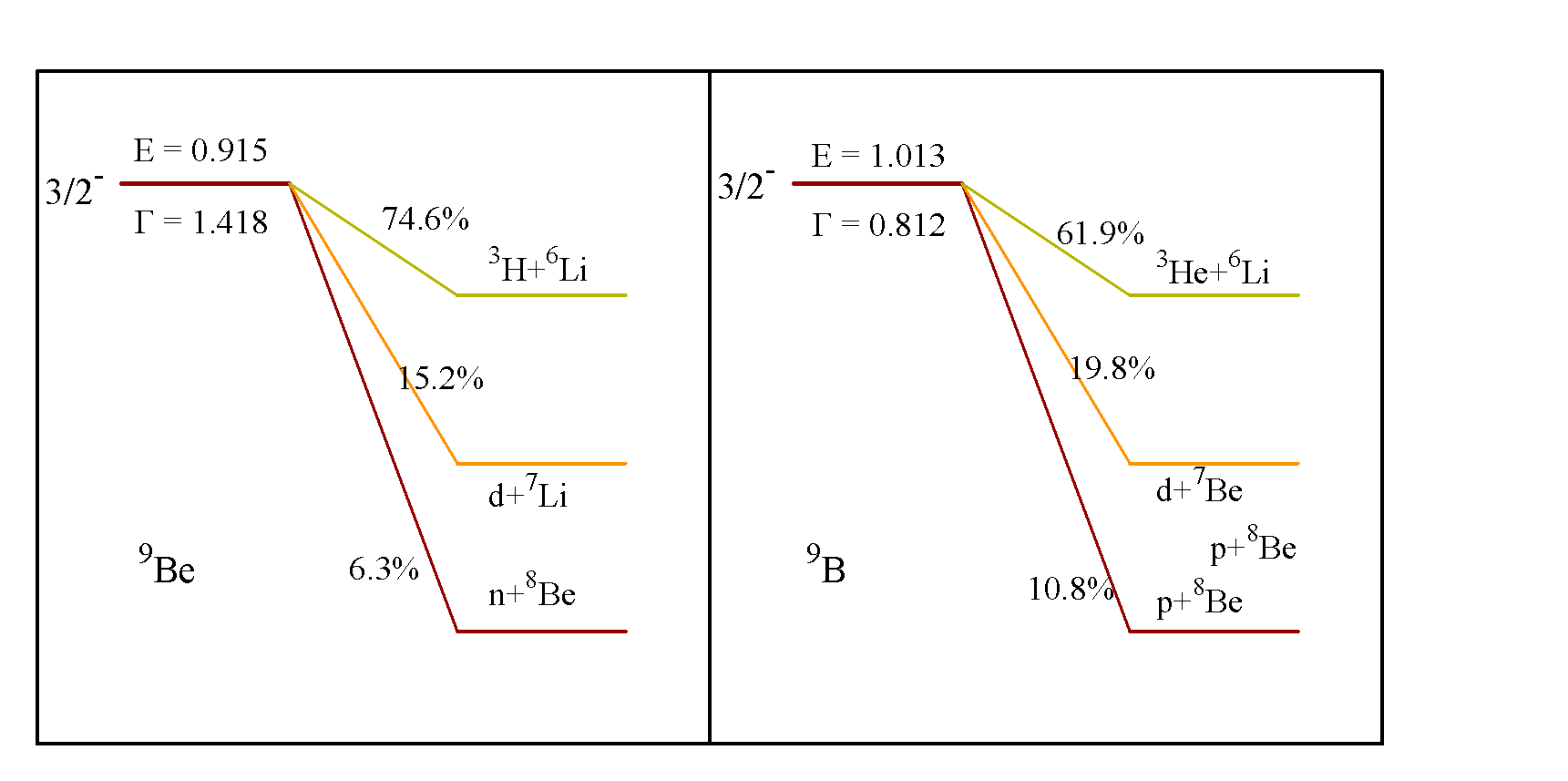}
\caption{Dominant decay channels of the 3/2$^{-}$ resonance states in $^{9}$Be
and $^{9}$B. Energies and widths are given in MeV, with energies measured relative to the $^7$Li$+d$ and $^7$Be$+d$ decay thresholds.}%
\label{Fig:DominDecayChR32M}%
\end{center}
\end{figure}

In Figs. \ref{Fig:DominDecayChR12M} and \ref{Fig:DominDecayChR32M} we
visualize results presented in Tables \ref{Tab:PartWidth9Be12M}\ and
\ref{Tab:PartWidth9B12M} show dominant decay channels. We display only
cumulative contribution of each binary channel. The 1/2$^{-}$ resonance states
decay with the largest probability through the channels $n$+$^{8}$Be and $p$%
+$^{8}$Be, respectively in $^{9}$Be and $^{9}$B. A significant probability of decay for
the 1/2$^{-}$ resonance states is determined by the  $d$+$^{7}$Li and
$d$+$^{7}$Be channels.%


\subsection{Effects of the Coulomb interaction}
\label{coulomb}

As we pointed out above, the differences between the structure of the 5/2$^{-}$
resonance states in $^{9}$Be and $^{9}$B primarily stem from the Coulomb
interaction. Let's elaborate on this statement. In our model, we employ two
three-cluster configurations $\alpha+\alpha+n$ and $\alpha+t+d$ for $^{9}$Be
and $\alpha+\alpha+p$ and $\alpha+^{3}$He$+d$ \ for $^{9}$B. We also use a common
oscillator length for all clusters and for both nuclei. 
Furthermore we apply the
same value of the exchange parameters $u$ for both nuclei. Consequently, the structure of
alpha particles and deuterons remains identical for $^{9}$Be and $^{9}$B. 
A neutron and proton are are considered structureless in our model. 
Therefore, the nuclear part of the interactions between cluster pairs within each three-cluster configurations  is the same. 
Besides, the nuclear part that determines the  coupling of binary channels $^{7}$Li+$d$ and
$^{6}$Li+$^{3}$H ($^{7}$Be+$d$ and $^{6}$Li+$^{3}$He) from the first
three-cluster configuration to binary channels $^{8}$Be+$n$ and $^{5}%
$He+$\alpha$ ($^{8}$Be+$p$ and $^{5}$Li+$\alpha$)  of the second
configuration, is also the same.   

However, when considering the Coulomb interaction and  projecting the
three-cluster configurations onto a set of binary channels, we encounter
different channels with varying structures of some two-cluster subsystems. 
This implies that, for instance, the structures (energies and wave functions) of $^{7}$Li and $^{7}$Be are distinct, and their interactions with a deuteron differ as well. The same holds true for the clusters $^{5}$He and $^{5}$Li, along with their interactions with an alpha particle.  Thus the Coulomb interaction and difference in the internal
structure of clusters $^{7}$Li and $^{7}$Be, $^{5}$He and $^{5}$Li are totally
responsible for the dissimilarity of the spectra of resonance states in mirror
nuclei $^{9}$Be and $^{9}$B.

The impact of the Coulomb interaction on low-energy resonance states in $^{9}$Be
and $^{9}$B has previously been studied in Ref. \cite{2020NuPhA.99621692D}. These investigations were conducted within a microscopic three-cluster model, which was originally developed in Ref. \cite{2001PhRvC..63c4606V}. This model utilizes hyperspherical harmonics to describe three-cluster continuum and
to identify resonance states embedded in the continuum. 
It was established in Ref. \cite{2001PhRvC..63c4606V} that strong effects of the Coulomb interaction are observed for bound and very narrow resonance states in $^{9}$Be and $^{9}$B, as these states possess the most compact three-cluster configurations. The minor effects were detected in wide resonance states of these nuclei with highly dispersed three-cluster configurations.  

As resonance states have two key parameters, namely, energy $E$ and width $\Gamma$, it was 
suggested in Ref. \cite{2020NuPhA.99621692D} to use a two-dimensional $E$ and
$\Gamma$ plane for analyzing the  effects of the Coulomb interaction. 
In Ref.
\cite{2020NuPhA.99621692D},  four possible scenarios of motion of resonance
states in this plane due to a stronger Coulomb interaction were discussed and only one
scenario was observed in the in mirror nuclei $^{9}$Be and $^{9}$B.  In this particular scenario, the energy and width of the resonance state increase in the nucleus with a stronger Coulomb
interaction compared to the corresponding state in the nucleus with a smaller
Coulomb interaction.

\begin{figure}
\begin{center}
\includegraphics[width=\textwidth]{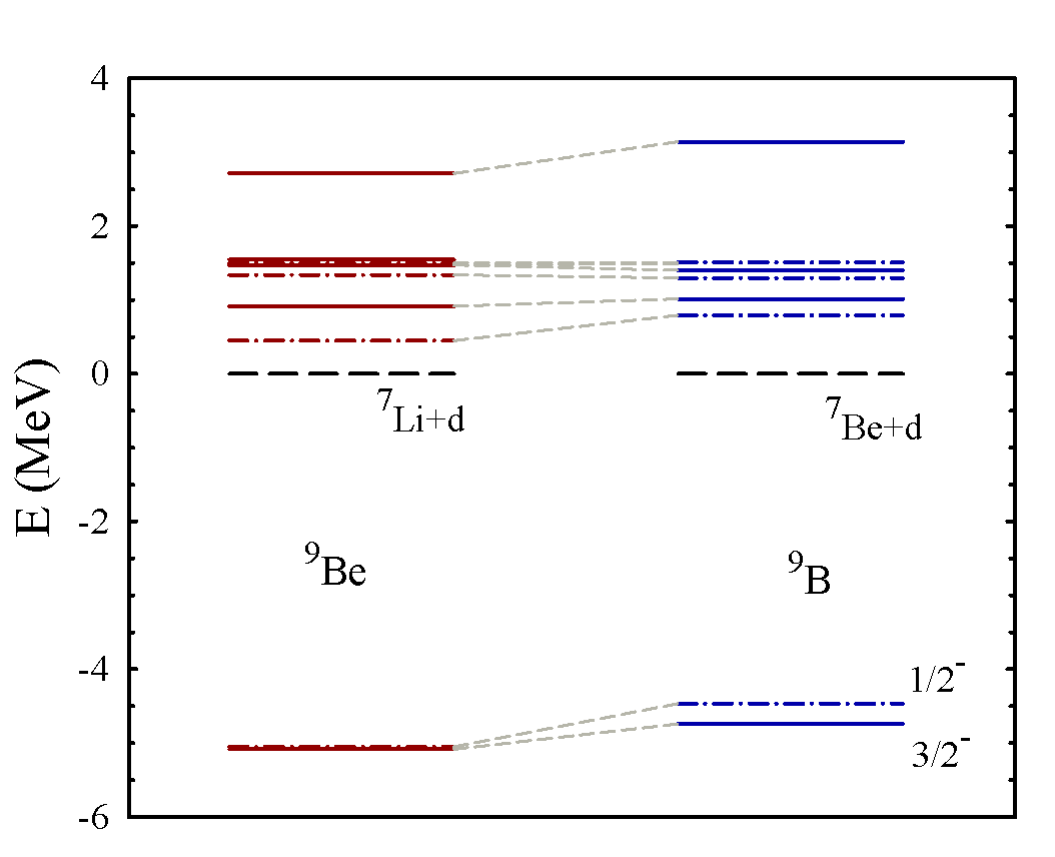}
\caption{Relative positions of negative parity resonance states in $^{9}$Be and $^{9}$B. Solid lines represent the 3/2$^{-}$ resonance states, and dash-dotted lines show the 1/2$^{-}$ resonance states.}%
\label{Fig:EffectsCoulm1}%
\end{center}
\end{figure}

\begin{figure}
\begin{center}
\includegraphics[width=\textwidth]{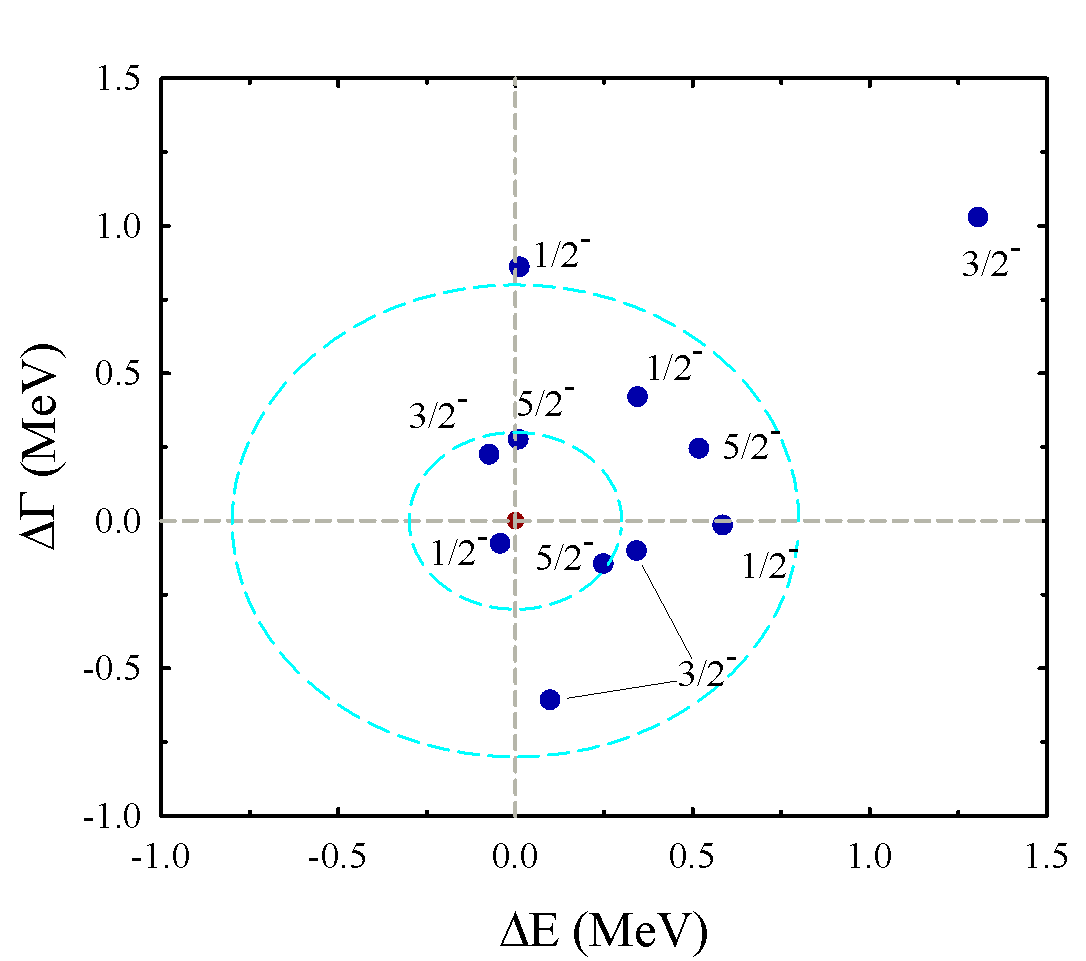}
\caption{Effects of the Coulomb interaction on  energies and widths of
negative parity resonance states in $^{9}$B (a red dot) and $^{9}$Be (blue dots).}%
\label{Fig:EffectsCoulm2}%
\end{center}
\end{figure}

In Figs. \ref{Fig:EffectsCoulm1} and \ref{Fig:EffectsCoulm2}, we illustrate the impact of Coulomb interaction on negative parity resonance states in $^{9}$Be and $^{9}$B. 
Fig. \ref{Fig:EffectsCoulm1} presents the traditional representation of Coulomb interaction effects, whereas Fig. \ref{Fig:EffectsCoulm2} introduces a novel approach. The first method demonstrates alterations in the energy of resonance states, while the second approach reveals the combined influence of the Coulomb interaction on both energy and width.

In Fig. \ref{Fig:EffectsCoulm2}, we have positioned all resonance states of $^{9}$B at the origin of the plane, marked by a red dot. The corresponding resonance states of $^{9}$Be are represented by blue points for comparison. The results presented in Fig. \ref{Fig:EffectsCoulm2} indicate that,  within the present model, the second, third, and fourth scenarios are realized in the mirror nuclei $^{9}$Be and $^{9}$B. The second scenario is related to an increase in energy and a decrease in width, the third scenario is realized when energy decreases and width increases, and the fourth scenario is realized when both energy and width decrease. 

Similarly to the low-energy resonance states (see Fig. 4 in Ref. \cite{2020NuPhA.99621692D}), the high-energy resonance states
in $^{9}$Be and $^{9}$B reveal significant, minor, and intermediate effects of the
Coulomb interaction. When the shift of resonance state (details of the definition
can be found in Ref. \cite{2020NuPhA.99621692D})%
\begin{eqnarray*}
R=\sqrt{\left(  \Delta E\right)  ^{2}+\left(  \Delta\Gamma\right)  ^{2}}%
\end{eqnarray*}
on the $E$ and $\Gamma$ plane is $R>0.8$ MeV, this corresponds to significant effects;
when $0.3$ MeV$<R<$0.8 MeV, these are intermediate effects of the Coulomb forces, and minor
effects are observed when $R<0.3$ MeV. In Fig. \ref{Fig:EffectsCoulm2}, two dashed circles mark the boundaries between regions of minor, intermediate and significant effects of the Coulomb interaction. %


\subsection{General features of resonant spectra in $^{9}$B and $^{9}$Be}
\label{general}
In Table \ref{Tab:SpectrRS9Bevs9B}, we present a comparative analysis of resonance state spectra in $^{9}$Be and $^{9}$B as derived from our model. A total of 18 resonance states were detected in each nucleus. Notably, we did not observe the 3/2$^{+}$ and 5/2$^{+}$ resonance states in either case. There are only a few positive parity states, with three resonances found in $^{9}$B and four in $^{9}$Be. However, a substantial number of negative parity states were identified, comprising 15 in $^{9}$B and 16 in $^{9}$Be. Among these, four resonance states result from the $J^{\pi}$ =3/2$^{-}$ channels in $^{9}$B, and a similar number arises from the $J^{\pi}$ =5/2$^{-}$ channels in $^{9}$Be.

The model used generates a very few narrow resonance states. Four resonance states in both $^{9}$B and $^{9}$Be have widths below 300 keV. All of them lie below the $^{7}$Be+$d$ and $^{7}$Li+$d$ thresholds, respectively. The narrowest resonance state in $^{9}$B above the $^{7}$Be+$d$ threshold is 0.812 MeV wide, while in $^{9}$Be, the most narrow resonance above the $^{7}$Li+$d$ threshold has a width of 0.680 MeV. These resonance states are unique in that they are relatively distant from their respective thresholds. Specifically, the lowest energy of resonance states in $^{9}$B is 0.793 MeV, and in $^{9}$Be, it is 0.448 MeV.

\begin{table}[htbp] \centering
\caption{Calculated spectrum of resonance states in $^9$B and  $^9$Be. Energies are measured relative to the $^7$Li$+d$ and $^7$Be$+d$ decay thresholds.}%
\begin{ruledtabular} 
\begin{tabular}
[c]{cccccc}
\multicolumn{3}{c}{$^{9}$Be} & \multicolumn{3}{c}{$^{9}$B}\\\hline
$J^{\pi}$ & $E$, MeV & $\Gamma$, MeV & $J^{\pi}$ & $E$, MeV & $\Gamma$,
MeV\\\hline
$\frac{1}{2}^{+}$ & -5.204 & 1.167 & $\frac{1}{2}^{+}$ & -6.424 &
1.294\\
$\frac{3}{2}^{-}$ & -5.084 & 0.183 & $\frac{3}{2}^{-}$ & -4.742 &
0.083\\
$\frac{1}{2}^{-}$ & -5.052 & 0.032 & $\frac{1}{2}^{-}$ & -4.467 &
0.018\\
$\frac{5}{2}^{-}$ & -4.413 & 0.245 & $\frac{5}{2}^{-}$ & -4.165 &
0.100\\
$\frac{7}{2}^{-}$ & -1.254 & 0.222 & $\frac{7}{2}^{-}$ & -0.300 &
0.116\\
$\frac{1}{2}^{-}$ & 0.448 & 0.680 & $\frac{1}{2}^{-}$ & 0.793 & 1.006\\
$\frac{3}{2}^{-}$ & 0.915 & 1.418 & $\frac{3}{2}^{-}$ & 1.013 & 0.812\\
$\frac{1}{2}^{-}$ & 1.339 & 1.101 & $\frac{1}{2}^{+}$ & 1.053 & 1.867\\
$\frac{3}{2}^{-}$ & 1.474 & 1.610 & $\frac{1}{2}^{-}$ & 1.296 & 1.025\\
$\frac{5}{2}^{-}$ & 1.478 & 1.262 & $\frac{3}{2}^{-}$ & 1.400 & 1.836\\
$\frac{1}{2}^{-}$ & 1.498 & 0.974 & $\frac{5}{2}^{-}$ & 1.418 & 1.585\\
$\frac{1}{2}^{+}$ & 1.513 & 1.238 & $\frac{5}{2}^{-}$ & 1.486 & 1.539\\
$\frac{3}{2}^{-}$ & 1.547 & 1.987 & $\frac{1}{2}^{-}$ & 1.509 & 1.834\\
$\frac{7}{2}^{+}$ & 2.526 & 1.803 & $\frac{7}{2}^{+}$ & 2.894 & 2.377\\
$\frac{5}{2}^{-}$ & 2.609 & 1.735 & $\frac{5}{2}^{-}$ & 2.914 & 1.832\\
$\frac{3}{2}^{-}$ & 2.714 & 2.453 & $\frac{7}{2}^{-}$ & 2.989 & 2.516\\
$\frac{7}{2}^{-}$ & 2.864 & 2.203 & $\frac{3}{2}^{-}$ & 3.142 & 3.482\\
$\frac{7}{2}^{+}$ & 2.945 & 1.827 & $\frac{7}{2}^{-}$ & 3.209 & 2.608\\
\end{tabular}
\end{ruledtabular} 
\label{Tab:SpectrRS9Bevs9B}%
\end{table}%

\begin{figure}
\begin{center}
\includegraphics[width=\textwidth]{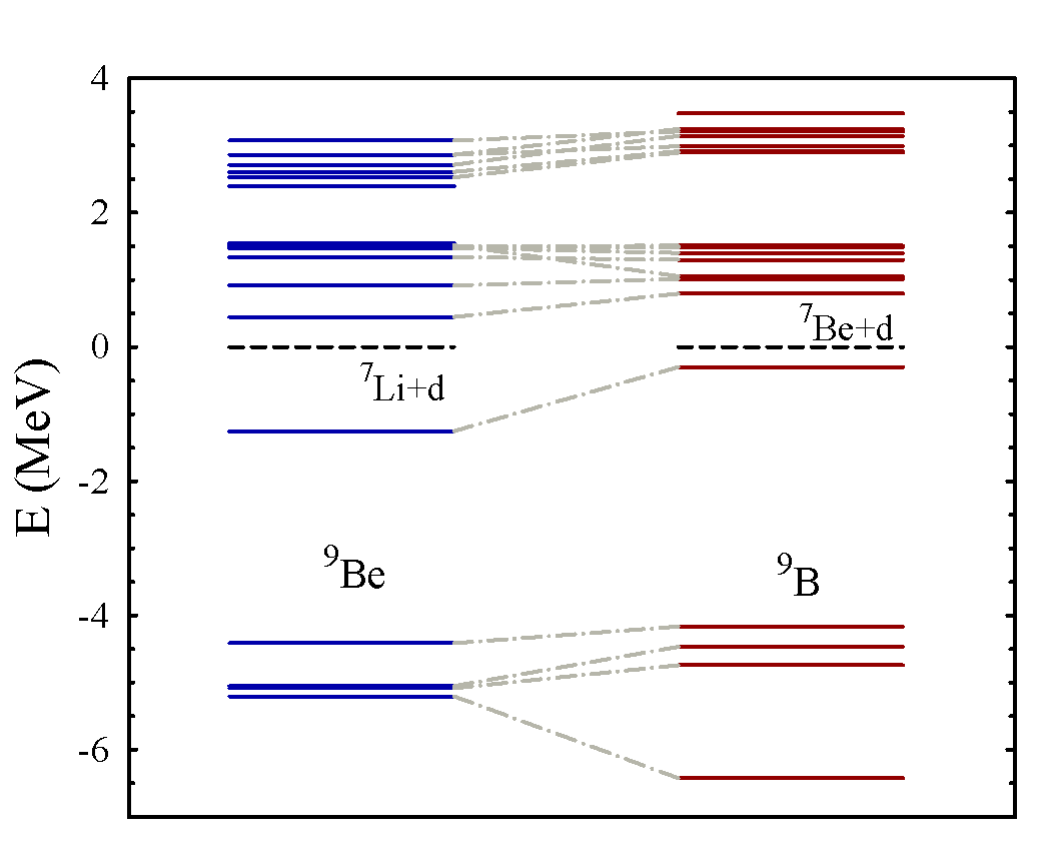}
\caption{Spectra of resonance states in $^{9}$Be and $^{9}$B obtained within
the present model. Grey dashed line connect resonance states with the same
values of the total angular momentum $J$ and parity $\pi$.}%
\label{Fig:SpectrRS9Bevs9B}%
\end{center}
\end{figure}

In Fig. \ref{Fig:SpectrRS9Bevs9B} we visualize spectrum of resonance states of
$^{9}$Be and $^{9}$B obtained within our model and displayed in Table
\ref{Tab:SpectrRS9Bevs9B}.  The dashed lines connect  resonance states
with the same quantum numbers $J$ and $\pi$.  Figure \ref{Fig:SpectrRS9Bevs9B} displays two distinct regions densely populated with resonance states. The first region encompasses energy levels ranging from 2.0 MeV to 3.0 MeV above the $^{7}$Li+$d$ threshold in $^{9}$Be and the $^{7}$Be +$d$ threshold in $^{9}$B. In this region, there are 6 resonance states in $^{9}$Be and 5 states in $^{9}$B. The second region contains 9 resonance states in $^{9}$Be and 8 resonance states in $^{9}$B, with energies falling between 0.4 and 2.0 MeV.
Additionally, a third region exists, featuring a smaller number of states, spanning from -6.5 to -4.0 MeV below the $^{7}$Li+$d$ and $^{7}$Be +$d$ thresholds. In this region, there are four resonance states each in $^{9}$Be and $^{9}$B.

\begin{figure}
\begin{center}
\includegraphics[width=\textwidth]{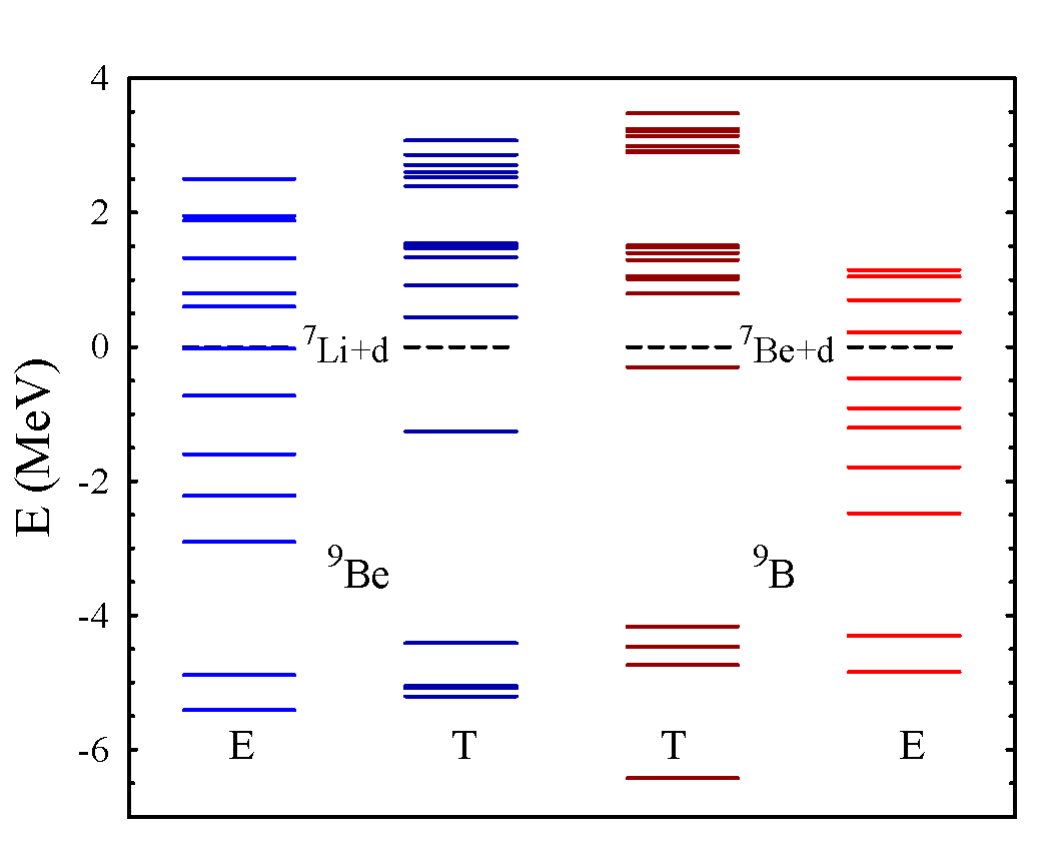}
\caption{Experimental (E) and theoretical (T) \ spectra  of high-lying
resonance states in $^{9}$Be and $^{9}$B.  Positions of the $^{7}$Li+$d$ threshold in $^{9}$Be and $^{7}$Be+$d$
threshold in $^{9}$B are indicated by the short-dashed lines.}%
\label{Fig:Spectr9Be9BTheorExp}%
\end{center}
\end{figure}

Fig. \ref{Fig:Spectr9Be9BTheorExp} compares resonance state parameters in $^{9}$B and $^{9}$Be from our many-channel model with experimental data \cite{2004NuPhA.745..155T}. We show 13 resonance states in $^{9}$Be and 11 in $^{9}$B, with $J$ and $\pi$ experimentally known for half of them. When matching experimental and theoretical spectra, an evident gap appears in the experimental energy spectrum below the $^{7}$Li+d and $^{7}$Be+d thresholds, within the energy range approximately from -5 to -3 MeV. In our calculations, this gap also exists but is broader, spanning over 3 MeV. However, the theoretical predictions for the resonances near the $^{7}$Li+d and $^{7}$Be+d thresholds, which could potentially account for the resonance behavior of the astrophysical S-factor, align reasonably well with the experimental data.

To conclude this section, we will summarize the main results derived from the analysis of resonance state properties. In both $^{9}$Be and $^{9}$B, strong coupling between binary channels is evident. The presence of $^{6}$Li+$^{3}$H channel  in $^{9}$Be and $^{6}$Li+$^{3}$He in $^{9}$B significantly impacts the bound and resonance states initially identified in the single-channel approach ($^{7}$Li+$d$ in $^{9}$Be and $^{7}$Be+$d$ in $^{9}$B). Importantly, both binary channels originate from the same three-cluster configuration ($\alpha$+$^{3}$H+$d$ in $^{9}$Be and $\alpha$+$^{3}$He+$d$ in $^{9}$B), with the second channel having a higher threshold energy (1.23 MeV above the first threshold in $^{9}$Be and 0.38 MeV in $^{9}$B). Effects of two other channels ($^{8}$Be+$n$ and $^{5}%
$He+$\alpha$ generated by three-cluster configuration $\alpha$+$\alpha$+$n$ in
$^{9}$Be and $^{8}$Be+$p$ and $^{5}$Li+$\alpha$ generated by three-cluster
configuration $\alpha$+$\alpha$+$p$ in $^{9}$B) are even stronger and
devastating. These two channels, when  coupled with the  first two channels,
dramatically change the energies and widths of resonance states, and may even dissolve
some of the bound and resonance states detected in two-channel approximations.

It was also shown that  the channels $^{8}$Be+$n$ and
$^{5}$He+$\alpha$  in $^{9}$Be, and $^{8}$Be+$p$ and $^{5}$Li+$\alpha$  in
$^{9}$B convert bound states produced by two other channels ($^{7}$Li+$d$ and
$^{6}$Li+$^{3}$H  in $^{9}$Be, and $^{7}$Be +$d$  and $^{6}$Li+$^{3}$He in
$^{9}$B) into Feshbach-type resonance states. 

Furthermore, in the current model, the effective
interaction between clusters  $^{7}$Li and $d$ in $^{9}$Be, as well as
$^{7}$Be  and $d$ in $^{9}$B, is fairly weak,  characterized by a shallow potential well. Consequently, it cannot produce narrow resonance states near their respective thresholds. Indeed, all resonance states in $^{9}$Be lie above 1.2 MeV
relative to the $^{7}$Li+$d$ threshold,  and in $^{9}$B, they are situated more than 0.8 MeV above the $^{7}$Be+$d$ threshold. 
As was shown in Table
\ref{Tab:SpectrRS9Bevs9B}, the smallest width of such resonances is 0.974
MeV in $^{9}$Be and 0.917 MeV in $^{9}$B. All positive-energy resonance states reside above the $^{6}$Li+$^{3}$H threshold in $^{9}$Be and the $^{6}$Li+$^{3}$He threshold in $^{9}$B. 

\subsection{S-factors}
\label{sfactor}
In this section we consider astrophysical reactions induced by the interaction
of deuteron with $^{7}$Li\ in $^{9}$Be%

\begin{equation}
d+^{7}\text{Li}=\left\{
\begin{array}
[c]{c}%
^{8}\text{Be}(0^{+})+n\\
^{8}\text{Be}(2^{+})+n\\
^{5}\text{He}(3/2^{-})+\alpha\\
^{5}\text{He}(1/2^{-})+\alpha
\end{array}
\right.  \label{eq:R01}%
\end{equation}

and by deuteron with $^{7}$Be in $^{9}$B%
\begin{equation}
d+^{7}\text{Be}=\left\{
\begin{array}[c]{c}
^{8}\text{Be}(0^{+})+p\\
^{8}\text{Be}(2^{+})+p\\
^{5}\text{Li}(3/2^{-})+\alpha\\
^{5}\text{Li}(1/2^{-})+\alpha
\end{array}
\right.  \label{eq:R02}
\end{equation}
Actually, all reactions under consideration involve three clusters
in the exit channels. However, we treat them as two-cluster reactions in both
the entrance and exit channels. Such a theoretical approach to these reactions
aligns with experimental investigations.

Astrophysical $S$ factors of reactions (\ref{eq:R01}) and (\ref{eq:R02})  are
studied in the energy range $0\leq E\leq2$ MeV. Special attention is paid to
the low-energy range which was dominant at the early stage of our Universe. 
We consider Gamow windows for entrance channels $d+^{7}$Li and $d+^{7}$Be at
a temperature $T_{9}$ =0.8 GK, as previously explored in
Refs. \cite{2005ApJ...630L.105A} and \cite{2019PhRvL.122r2701R}. 
This temperature uniquely defines the center and width of the Gamow peak energy \cite{1999NuPhA.656....3A} for the reactions induced by $d+^{7}$Be ($E_{0}$ = 307 keV, $\Delta E_{0}$ = 336 keV) and $d+^{7}$Li ($E_{0}$ = 253 keV, $\Delta E_{0}$ = 305 keV) interactions.

We calculate astrophysical $S$ factors for reactions (\ref{eq:R01}) and (\ref{eq:R02}) across a range of total angular momentum $J$ from 1/2 to 7/2, including both positive and negative parity states.

\subsubsection{Total and partial $S$ factors in $^9$B}

\begin{figure}
\begin{center}
\includegraphics[width=\textwidth]{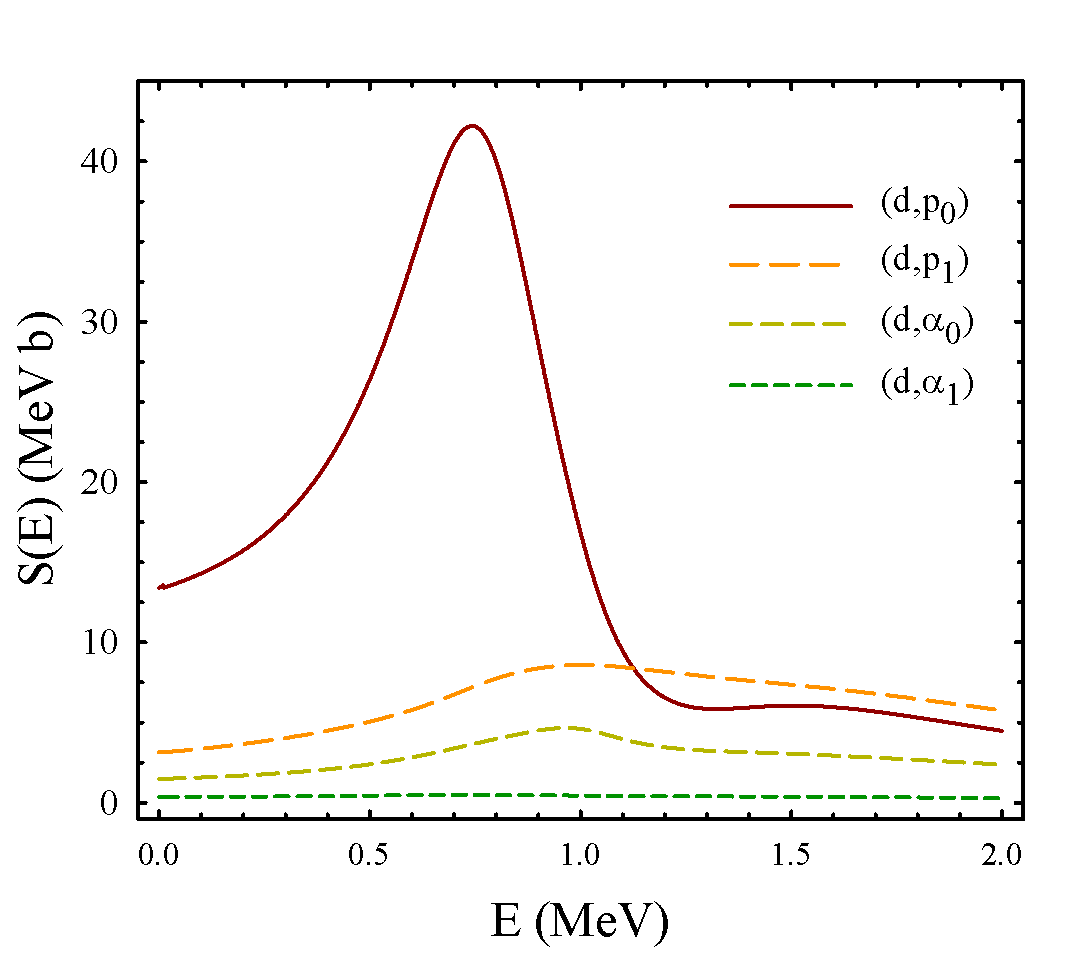}
\caption{Total astrophysical $S$ factors of four reactions initiated by the $^{7}$Be+$d$
interaction.}%
\label{Fig:Sfactors9BComp}%
\end{center}
\end{figure}

We begin by examining the total astrophysical $S$ factors of four reactions
(\ref{eq:R01}) in $^{9}$B.  Fig. \ref{Fig:Sfactors9BComp} indicates a strong hierarchy among the reactions in a small energy region. The reaction
$d+^{7}$Be$=^{8}$Be$(0^{+})+p$, denoted  as $\left(  d,p_{0}\right)$,  
significantly dominates in the energy range 0$\leq E\leq$1 MeV. In contrast, the second reaction, $d+^{7}$Be$=^{8}$Be$(2^{+})+p$ (denoted as $\left(  d,p_{1}\right)$), slightly dominates when the energy in the entrance channel exceeds 1.1 MeV. 
The astrophysical $S$ factors for the reactions $d+^{7}$Be$=^{5}$Li$(3/2^{-})+\alpha$ ($d,\alpha_{0}$) and $d+^{7}$Be$=^{5}$Li$(1/2^{-})+\alpha$ ($d,\alpha_{1}$) are considerably smaller than the $S$ factors of the two previous reactions across the entire energy range. 

\begin{figure}
\begin{center}
\includegraphics[width=\textwidth]{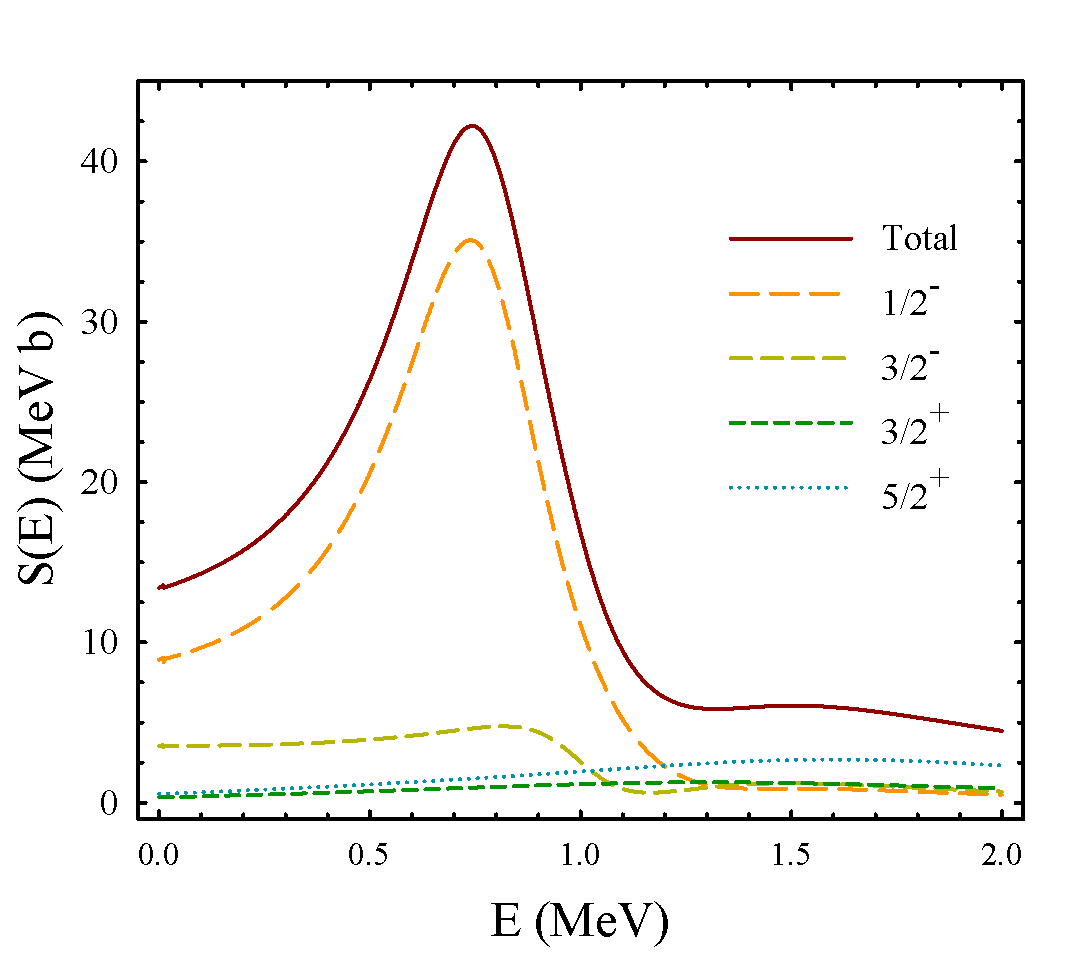}
\caption{The total and dominant partial $S$ factors of the reaction $^{7}$Be+$d$=$^{8}$Be($0^{+}$)+$p$}%
\label{Fig:Sfactor9BR1Dom}
\end{center}
\end{figure}

Astrophysical $S$ factors for the reaction $d+^{7}$Be$=^{8}$Be$(0^{+})+p$ has a significant peak at the
energy $E$=0.74 MeV. To understand origin of the peak, we analyse the total
and dominant partial astrophysical $S$ factors for the reaction $d+^{7}$Be$=^{8}$Be$(0^{+})+p$. In Fig. \ref{Fig:Sfactor9BR1Dom} we display only those partial  $S$ factors that make a noticeable contribution to the total $S$ factor for the reactions. Clearly, the partial $S$ factor for the 1/2$^{-}$ state dominates the energy range 0$\leq E\leq$1.1 MeV with a substantial peak.

Results presented in Table \ref{Tab:SpectrRS9Bevs9B} indicate that this peak is caused by the broad 1/2$^{-}$ resonance state,  with an energy of $E$=0.79 MeV  close to the peak energy $E$=0.74 MeV. It is surprising that\ such a wide resonance
state (total width $\Gamma$=1.0 MeV) creates such a significant peak in the
astrophysical $S$ factor. The small difference between the energy of the resonance
state and the energy of the $S$ factor peak suggests that the background
processes (or the nonresonant part) of the matrix elements of the
$S$ matrix) play important role and affect both parameters of resonance states
and astrophysical $S$ factors.

\begin{figure}
\begin{center}
\includegraphics[width=\textwidth]{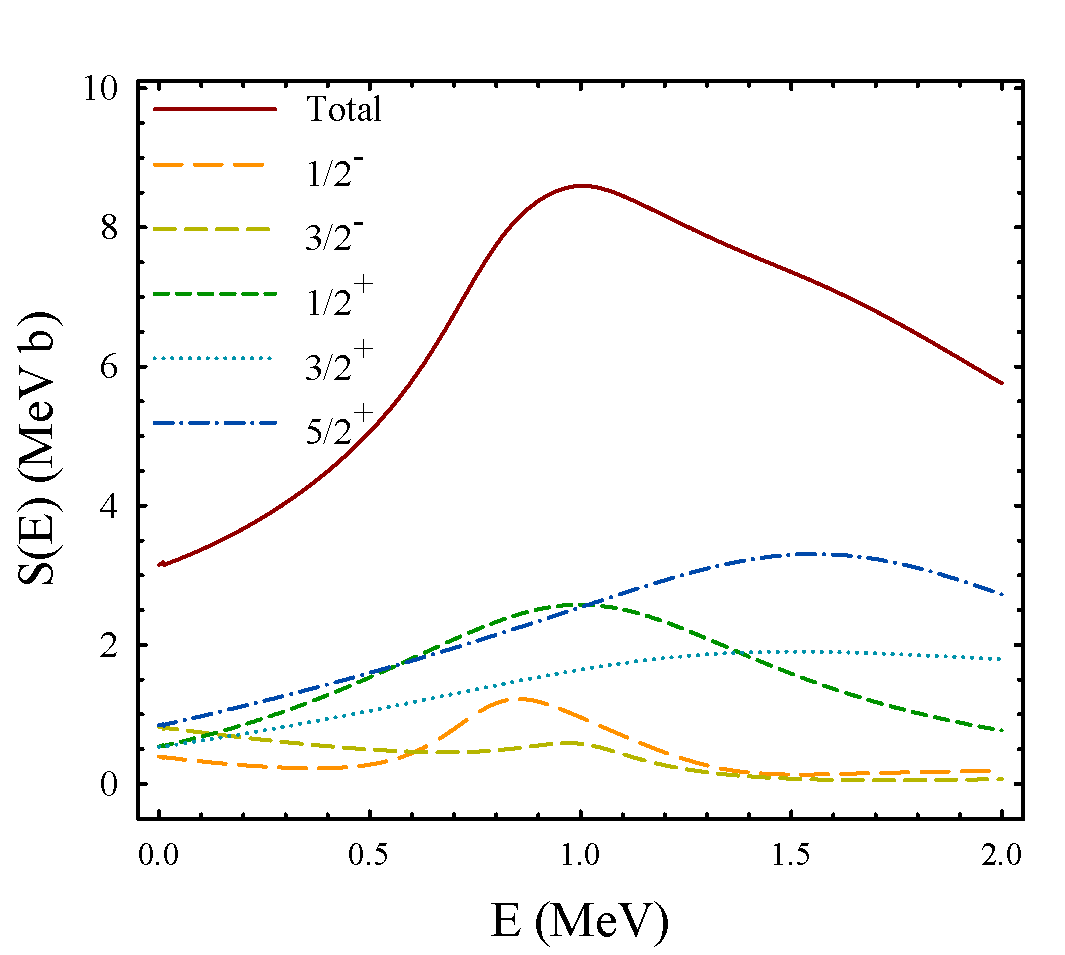}
\caption{The total and dominant partial astrophysical $S$ factors of the reaction
$d+^{7}$Be$=^{8}$Be$(2^{+})+p$}
\label{Fig:Sfactor9BR2Dom}%
\end{center}
\end{figure}

Unlike the first $\left(  d,p_{0}\right)  $ reaction, the second $\left(
d,p_{1}\right)  $ reaction lacks a strongly dominant state. The astrophysical
$S$ factor of the reaction (see Fig. \ref{Fig:Sfactor9BR2Dom}) is almost
uniformly spread over the states 1/2$^{+}$, 3/2$^{+},$ and 5/2$^{+}$. In contrast to the the first $\left(  d,p_{0}\right)$ reaction, the contribution of
negative parity states is very small.

\begin{figure}
\begin{center}
\includegraphics[width=\textwidth]{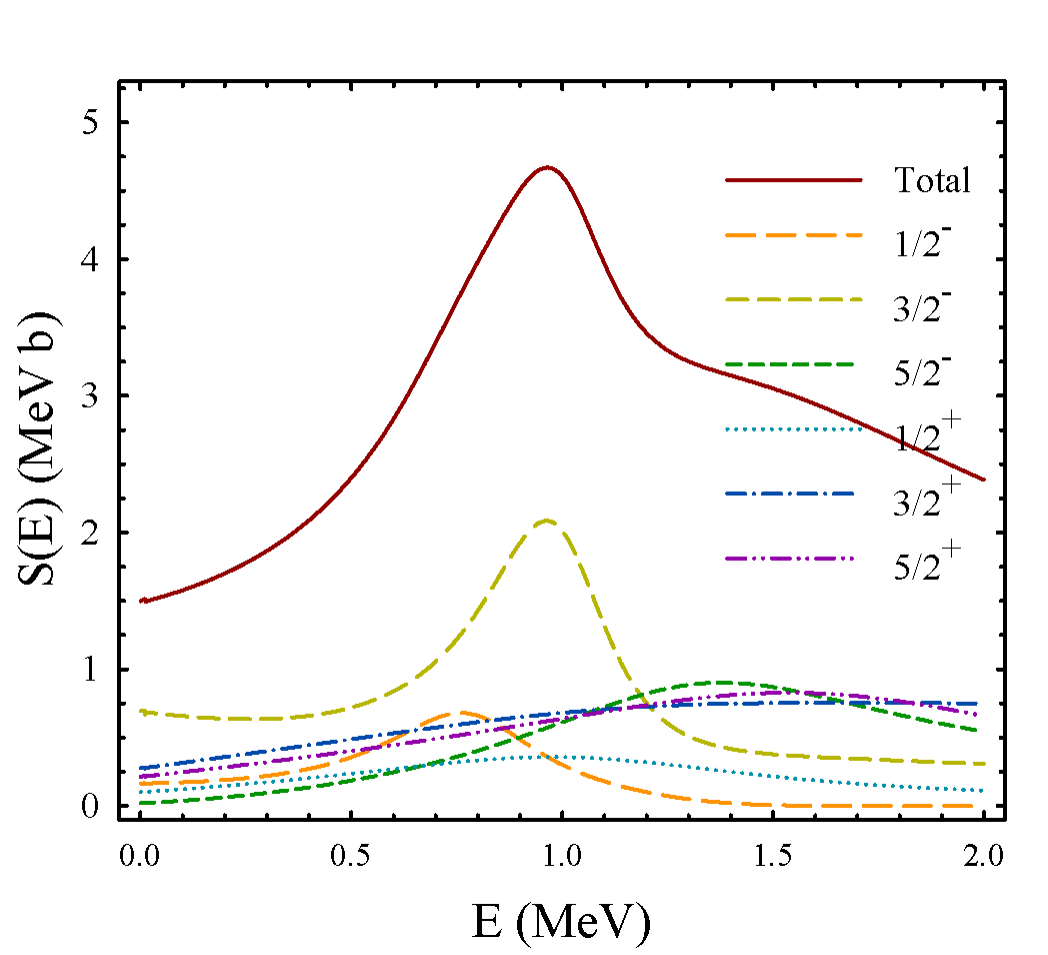}
\caption{The total and dominant partial $S$ factors of the reaction $d+^{7}$Be$=^{5}$Li$(3/2^{-})+\alpha$}
\label{Fig:Sfactor9BR3Dom}%
\end{center}
\end{figure}

In Fig. \ref{Fig:Sfactor9BR3Dom} we display the total and dominant partial $S$
factors of the reaction $\left(  d,\alpha_{0}\right)$. The total $S$ factor
exhibits a very broad peak centered at $E$=0.96 MeV, attributed to the
3/2$^{-}$ resonance state. The partial astrophysical $S$ factor associated with the
3/2$^{-}$ state dominates at  the energy range 0$\leq E\leq$1.15 MeV.

\subsubsection{Total and partial $S$ factors in $^9$Be}

\begin{figure}
\begin{center}
\includegraphics[width=\textwidth]{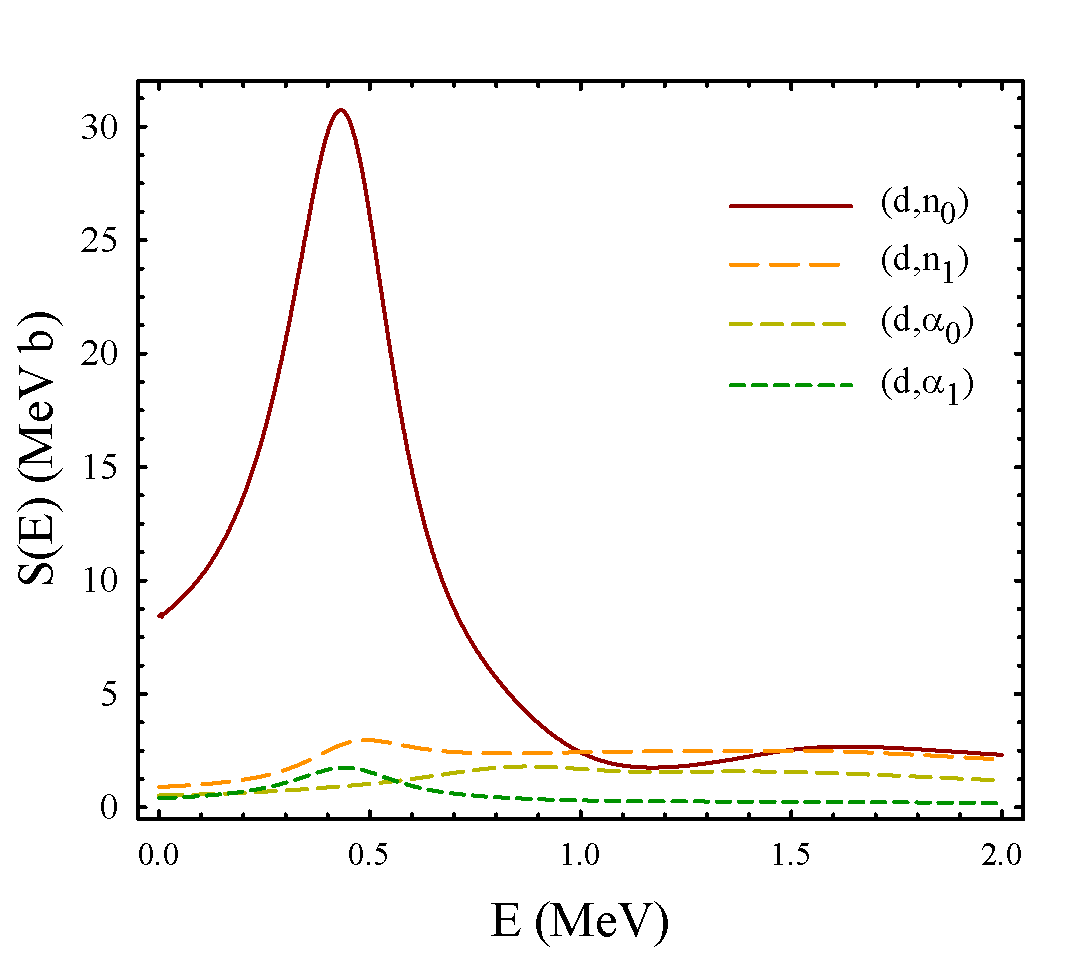}
\caption{Total $S$ factors of reactions induced by interaction of deuterons
with $^{7}$Li}
\label{Fig:Sfactors9BeComp}
\end{center}
\end{figure}

\begin{figure}
\begin{center}
\includegraphics[width=\textwidth]{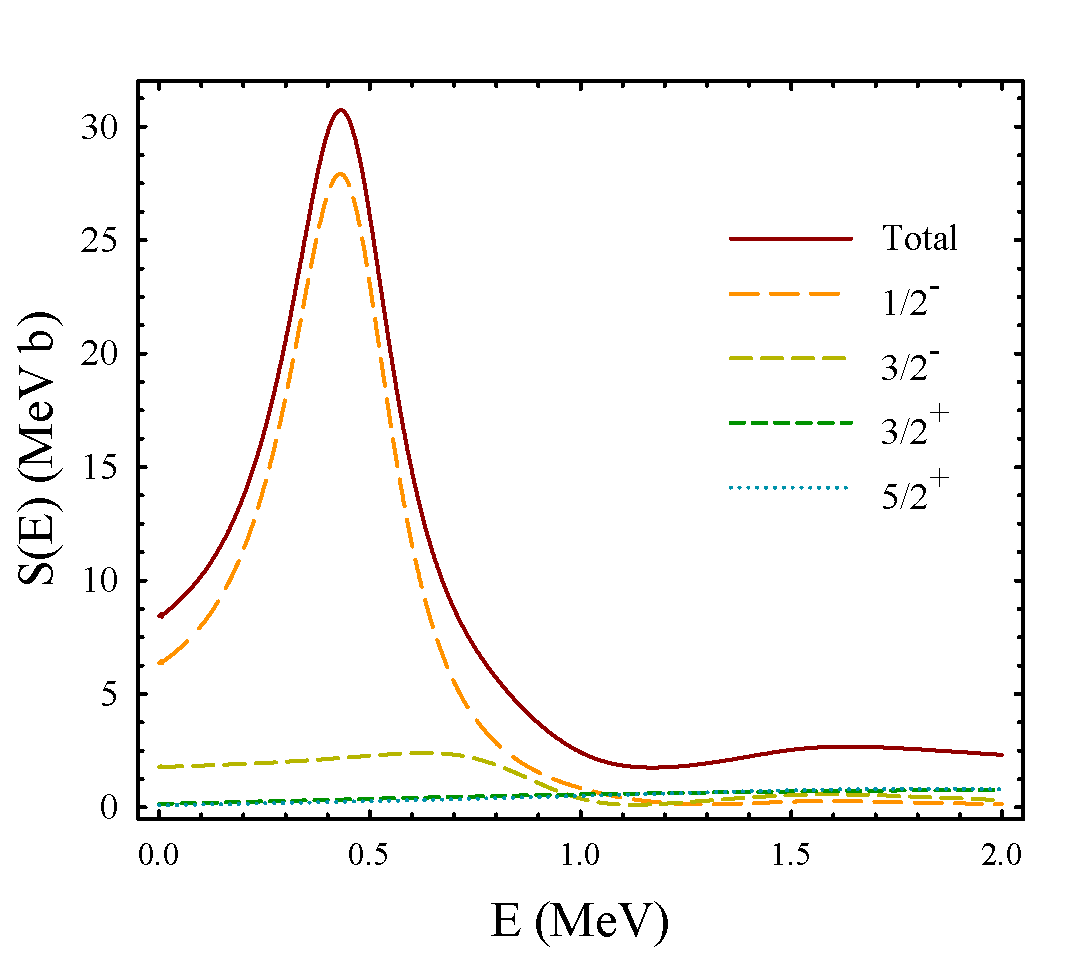}
\caption{The total and partial $S$ factors of the reaction $^{7}$Li+d=$^{8}%
$Be($0^{+}$)+$n$}%
\label{Fig:Sfactor9BeR1Dom}%
\end{center}
\end{figure}

The hierarchy of the four reactions in $^{9}$Be is similar to the one in
$^{9}$B. Indeed, as it shown in Fig. \ref{Fig:Sfactors9BeComp} and
Fig. \ref{Fig:Sfactors9BComp} the astrophysical $S$ factors for the reactions
$d+^{7}$Li$=^{8}$Be$(0^{+})+n$ and $d+^{7}$Be$=^{8}$Be$(0^{+})+p$  are
significantly larger than the $S$ factors for other reactions.  The $S$
factors for both reactions have huge peaks associated with the 1/2$^{-}$
resonance states. This is also demonstrated in Fig. \ref{Fig:Sfactor9BeR1Dom}
where the total and dominant partial $S$ factors of the reactions generated by
the $d+^{7}$Li collision are displayed. The energies of the peaks are slightly
different: $E_{0}$ = 0.45 MeV in $^9$Be and $E_{0}$ = 0.74 MeV in 9B.

\subsubsection{$S$-factors at Gamow peak energy}

\begin{figure}
\begin{center}
\includegraphics[width=\textwidth]{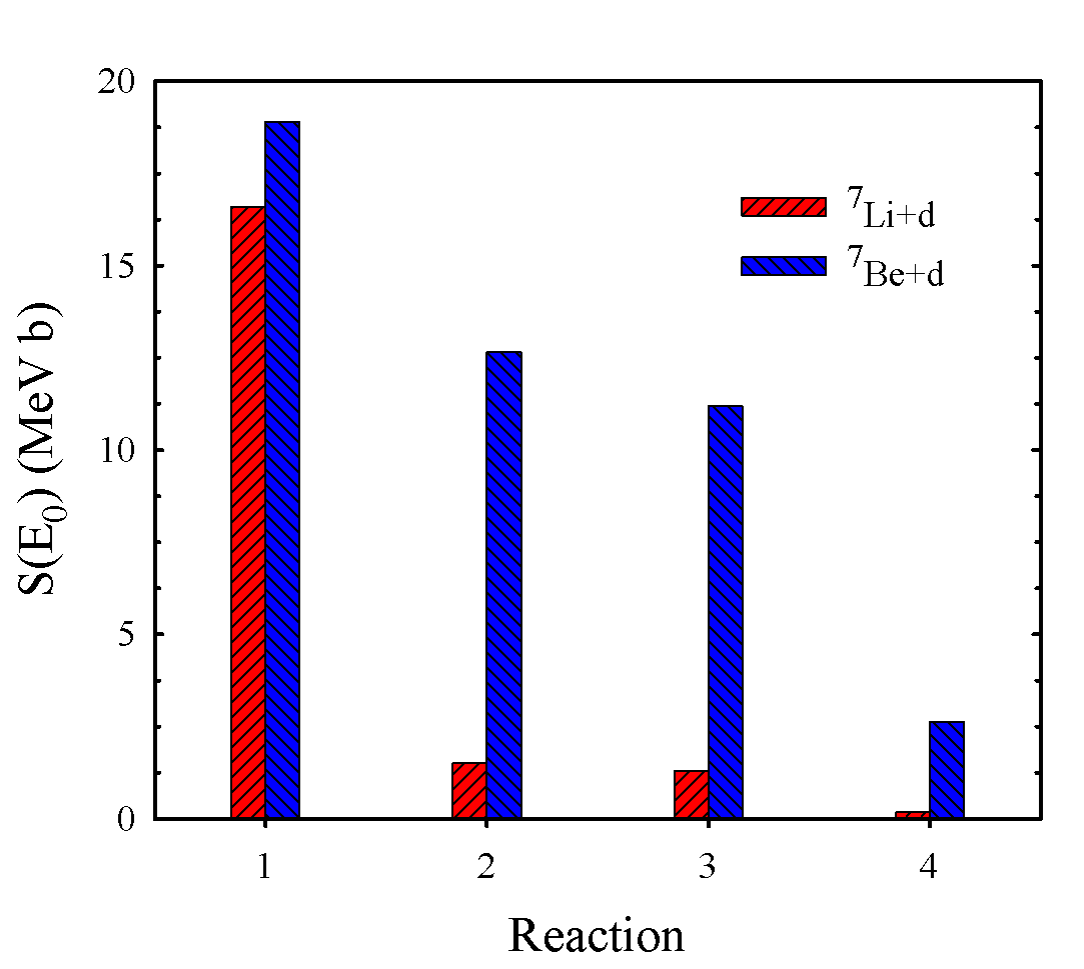}
\caption{Total astrophysical $S$ factor of the four reactions generated by
 the $^{7}$Li+d and $^{7}$Be+d collisions at the Gamow peak energies $E_{0}\left(  d+^{7}\text{Li}\right)  $ = 253 keV and $E_{0}\left(  d+^{7}\text{Be}\right)  $ =  307 keV. Exit channels correspond to the following reactions: 
$1 \Rightarrow$$^{8}$Be$(0^{+})+n$ ($^{8}$Be$(0^{+})+p$); $2 \Rightarrow$ $^{8}$Be$(2^{+})+n$ ($^{8}$Be$(2^{+})+p$); 
$3 \Rightarrow$ $^{5}$He$(3/2^{-})+\alpha$ ($^{5}$Li$(3/2^{-})+\alpha$), and $4 \Rightarrow$ $^{5}$He$(1/2^{-})+\alpha$ ($^{5}$Li$(1/2^{-})+\alpha$).}%
\label{Fig:Sfactors72E0}%
\end{center}
\end{figure}

Let us consider total astrophysical $S$ factors for reactions induced by the
collisions $d+^{7}$Li and $d+^{7}$Be at the corresponding  Gamow peak energies
$E_{0}\left(  d+^{7}\text{Li}\right)  $ = 253 keV and $E_{0}\left(  d+^{7}\text{Be}\right)  $ = 307 keV. In Fig. \ref{Fig:Sfactors72E0}, we present four reactions, where $Reaction=1$ corresponds to the exit channel $^{8}$Be$(0^{+})+n$ ($^{8}$Be$(0^{+})+p$), $Reaction=2$ to the exit channel $^{8}$Be$(2^{+})+n$ ($^{8}$Be$(2^{+})+p$), $Reaction=3$ to the exit channel $^{5}$He$(3/2^{-})+\alpha$ ($^{5}$Li$(3/2^{-})+\alpha$), and $Reaction=4$ represents the exit channel $^{5}$He$(1/2^{-})+\alpha$ ($^{5}$Li$(1/2^{-})+\alpha$). It is interesting that the $S$ factors of the
reactions $d+^{7}$Li=$^{8}$Be$(2^{+})+n$ and $d+^{7}$Be=$^{8}$Be$(2^{+})+p$
are the largest and close to each other at these specific energies. As for
other reactions, the $S$ factors of the reactions generated by the $d+^{7}$Be
interaction  are much larger than the $S$ factors of the corresponding
reactions generated by the $d+^{7}$Li interaction.

\begin{figure}
\begin{center}
\includegraphics[width=\textwidth]{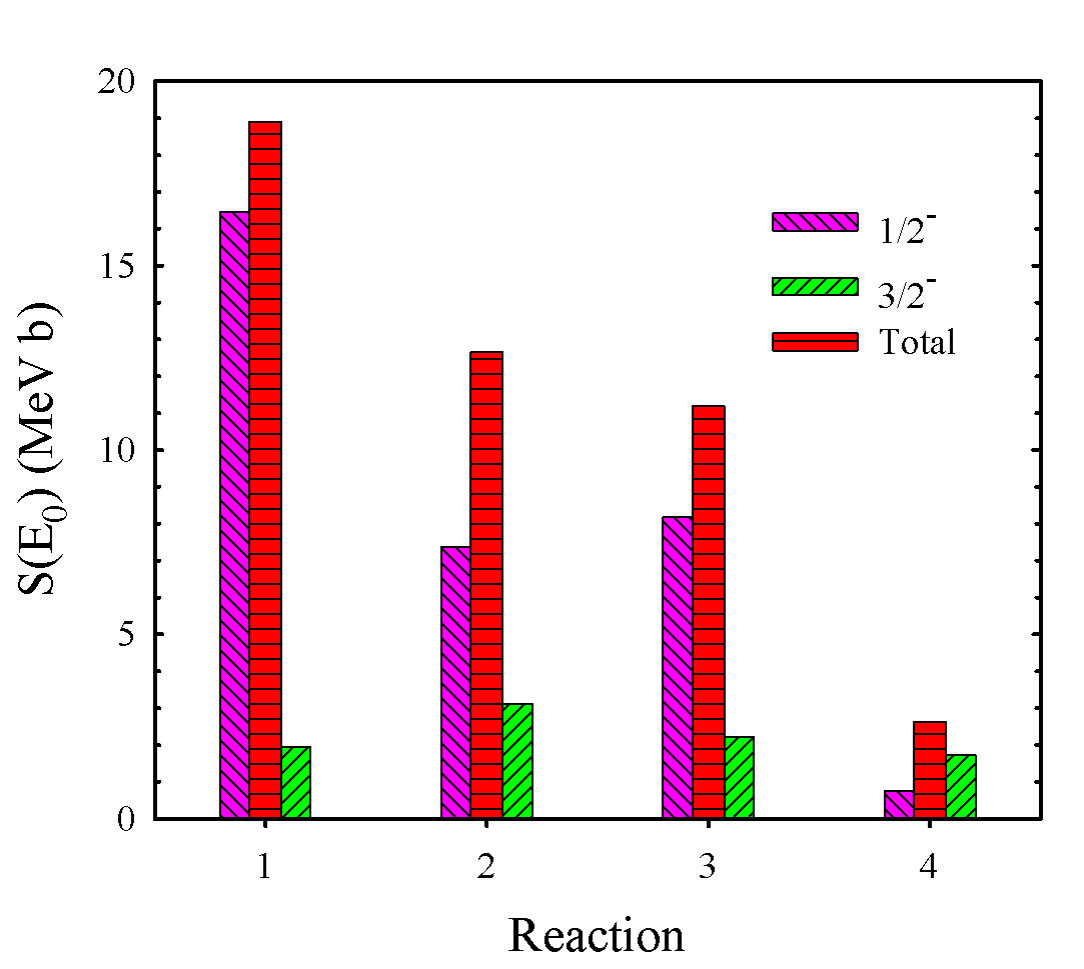}
\caption{Dominant states through which reactions induced by the $^{7}$Be+$d$
collisions proceeds at the Gamow peak energy $E_{0}\left(  d+^{7}\text{Be}\right)  $=307 keV.  Exit channels correspond to the following reactions: 
$1 \Rightarrow^{8}$Be$(0^{+})+p$; $2 \Rightarrow^{8}$Be$(2^{+})+p$; 
$3 \Rightarrow^{5}$Li$(3/2^{-})+\alpha$, and $4 \Rightarrow^{5}$Li$(1/2^{-})+\alpha$.}%
\label{Fig:Sfactors7BeDE0}%
\end{center}
\end{figure}

Fig. \ref{Fig:Sfactors7BeDE0} shows two dominant $J^{\pi}$ states  in
reactions induced by the $d+^{7}$Be collision: the 1/2$^{-}$ and
3/2$^{-}$ states. Head-on collision of a deuteron with the nucleus $^{7}$Be are
possible in these states, occurring when deuteron is moving
with zero orbital momentum relative to $^{7}$Be. The contributions of the
1/2$^{-}$ and 3/2$^{-}$ states are 83\%  for the second reaction $d+^{7}$Be$\rightarrow^{8}$Be$(2^{+})+p$, and exceeds 95\% for the remaining three reactions.

\subsubsection{Theory versus Experiment: $^9$B}

\begin{figure}
\begin{center}
\includegraphics[width=\textwidth]{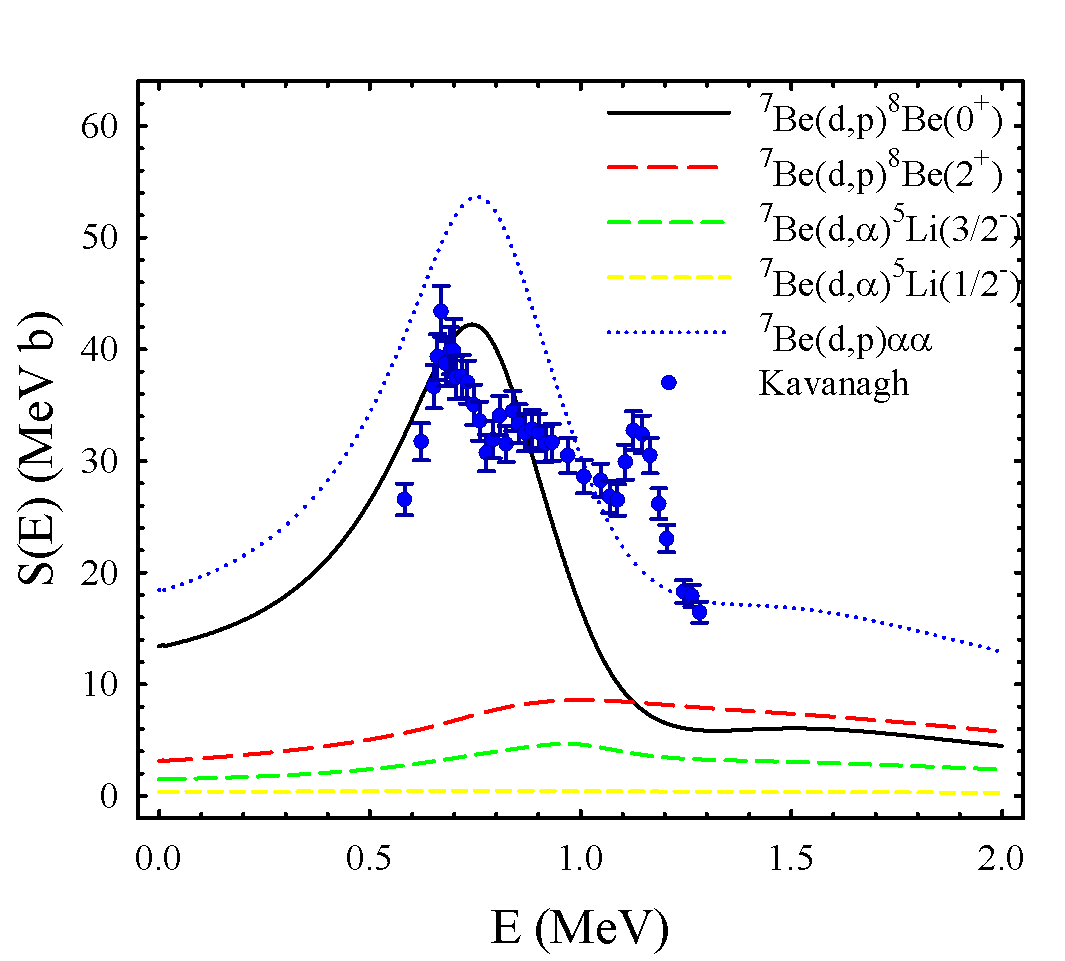}
\caption{Comparison of our results with the experimental data obtained
by Kavanagh \cite{1960NucPh..18..492K}}%
\label{Fig:Sfactors7BeDTvsE}%
\end{center}
\end{figure}

We compare results of our calculations of the astrophysical $S$ factor with
available experimental data. We start with reactions induced by the
interaction of deuterons with $^{7}$Be, as these experiments have a richer
history and provide more information.  In Fig \ref{Fig:Sfactors7BeDTvsE} we compare
our results with the experimental data obtained by Kavanagh in Ref.
\cite{1960NucPh..18..492K}. The total astrophysical $S$ factor of the reaction
$^{7}$Be($d$,$p$)$^{8}$Be(0$^{+}$) (solid line) obtained within the present
model reproduce fairly well experimental data around main peak located at
$E$=0.68 MeV. As we demonstrated above, the peak at $E$=0.73  MeV in our
model is associated with the 1/2$^{-}$ resonance state. Our model does not
reproduce the second peak at 1.15 MeV as there is no resonance state at this
energy in the dominant channels of the reactions. %

\begin{figure}
\begin{center}
\includegraphics[width=\textwidth]{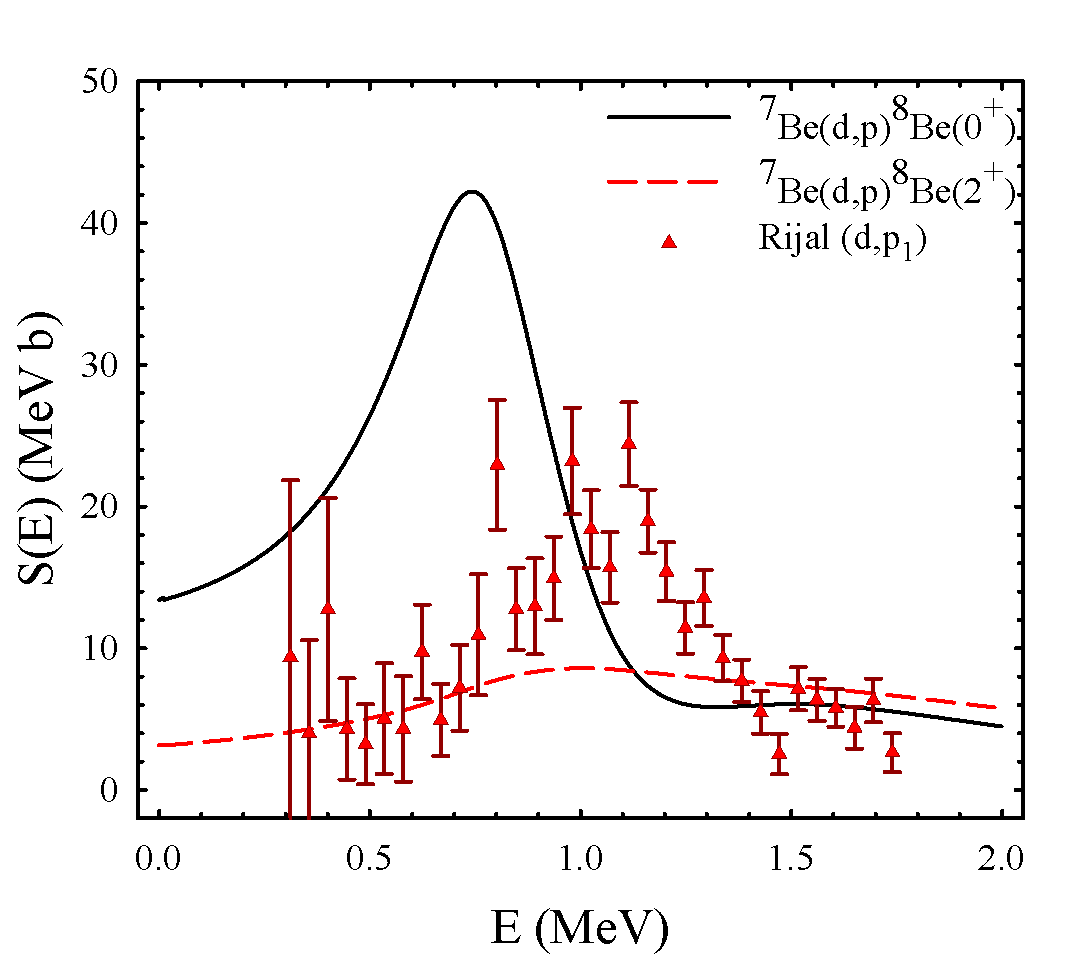}
\caption{Astrophysical $S$ factors of the reactions   $^{7}$Be($d$,$p$)$^{8}%
$Be(0$^{+}$) and $^{7}$Be($d$,$p$)$^{8}$Be(2$^{+}$) \ compared with
experimental data of Rijal et al. \cite{2019PhRvL.122r2701R}}.
\label{Fig:Sfactors7BeDP_TvsE}%
\end{center}
\end{figure}

In Fig. \ref{Fig:Sfactors7BeDP_TvsE}, the total astrophysical $S$ factors of
the reactions $^{7}$Be($d$,$p$)$^{8}$Be(0$^{+}$) and $^{7}$Be($d$,$p$)$^{8}%
$Be(2$^{+}$) are compared with the recent experimental data presented in Ref.
\cite{2019PhRvL.122r2701R}. Experimental data of Rijal et al in Ref.
\cite{2019PhRvL.122r2701R} obtained only for the second reaction $^{7}$%
Be($d$,$p$)$^{8}$Be(2$^{+}$). For the sake of complete analysis, we presented
$S$ factors of both reactions. At small energies 0.3$<E<$0.75 MeV our results
are very close to the experimental data. The same is correct for energy range
$E>$1.5 MeV. The theoretical $S$ factor has a peak at an energy close to
the experimental peak, although the height of the theoretical peak
is smaller than that of the experimental one. The significant peak in the astrophysical $S$ factor is generated by the reaction $^{7}$Be($d$,$p$)$^{8}$Be(0$^{+}$). It is worth noting that our calculations do not reveal any
narrow resonance state near the threshold $d+^{7}$Be — specifically, in the energy
range which is not accessible using modern experimental methods. This was
believed to hold promise for resolving the cosmological lithium problem.%

\subsubsection{Theory versus Experiment: $^9$Be}

To compare our results for the reactions induced by interaction of a deuteron 
with $^{7}$Li with experimental data, we relay on Refs.
\cite{2021ApJ...920..145H} and \cite{2006PhRvC..73a5801S}. In Ref.
\cite{2021ApJ...920..145H}, the experimental astrophysical $S$ factor of the
reaction $^{7}$Li($d$,$n$)$^{8}$Be is presented in a wide energy range
0$\leq E\leq1$ MeV. This $S$ factor includes results for two reactions $^{7}$Li($d$,$n_{0}$) and $^{7}$Li($d$,$n_{1}$). Meanwhile, in Ref. \cite{2006PhRvC..73a5801S} the astrophysical $S$ factors were determined for
reactions $^{7}$Li($d$,$n_{0}$) and $^{7}$Li($d$,$n_{1}$) separately. The
simple linear approximation has been used to present these $S$ factors in the low
energy range 0$\leq E<0.1$ MeV:%
\[
S(E)=400(\pm200)+0.62(\pm3.3)\cdot E
\]
for the first $^{7}$Li($d$,$n_{0}$),
\[
S(E)=5000(\pm1500)-37(\pm21)\cdot E
\]
for the reaction $^{7}$Li($d$,$n_{1}$) and then the total $S$ factor
\[
S(E)=5400(\pm1500)-37(\pm21)\cdot E.
\]
Here, astrophysical $S$ factors are given in keV$\cdot$b , with the energy measured in keV.

\begin{figure}
\begin{center}
\includegraphics[width=\textwidth]{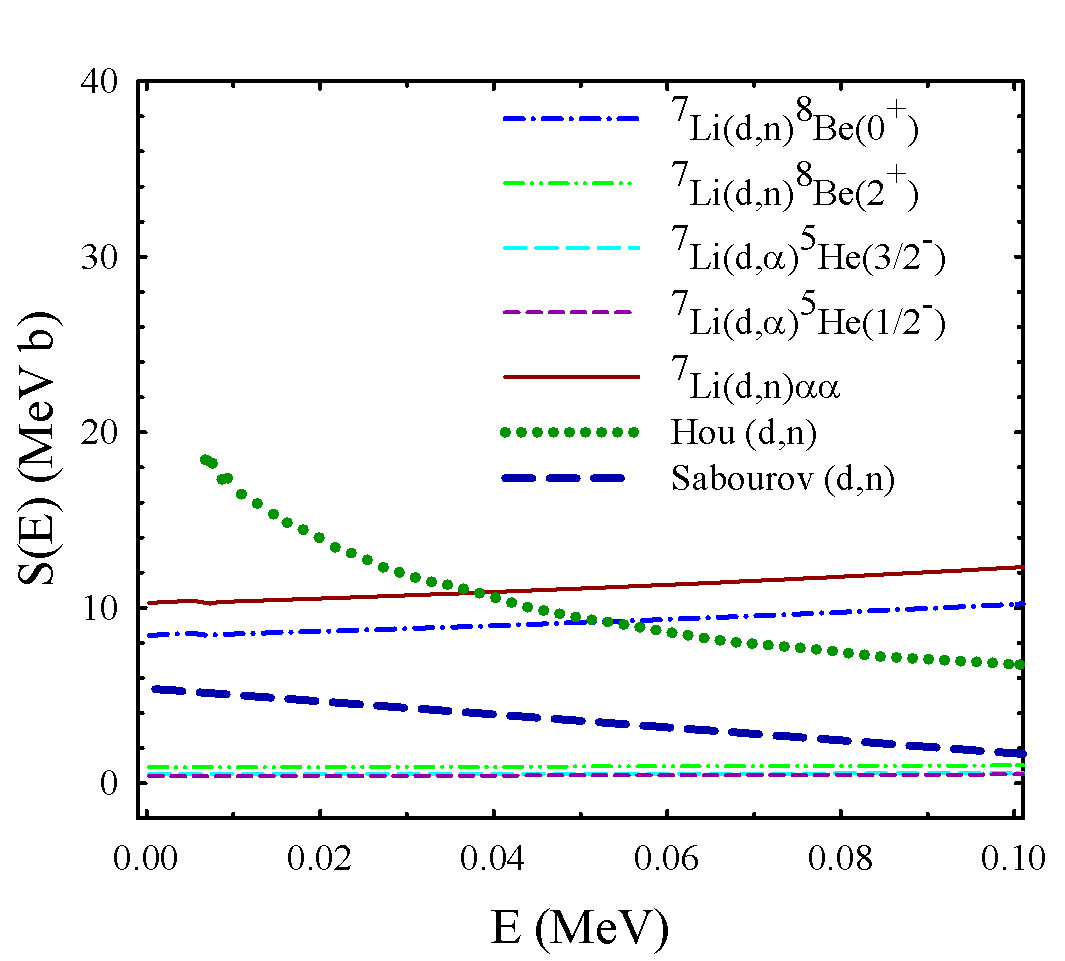}
\caption{Calculated astrophysical $S$ factor of the reactions induced by
low-energy scattering of deuteron on $^{7}$Li, and compared with the experimental data
from \cite{2021ApJ...920..145H} (Hou) and \cite{2006PhRvC..73a5801S} (Sabourov)}
\label{Fig:Sfactors7LiDNTvsE}
\end{center}
\end{figure}

In Fig. \ref{Fig:Sfactors7LiDNTvsE} we compare our results, which include
astrophysical $S$ factors of four reactions and total $S$ factor denoted as
$^{7}$Li($d$,$n$)$\alpha\alpha$, with experimental data Hou et al
\cite{2021ApJ...920..145H} and Sabourov et al \cite{2006PhRvC..73a5801S}. Here we consider the later data for the total $S$ factor only. In Fig. \ref{Fig:Sfactors7LiDNTvsE}, we have confined our analysis to the energy range where the linear approximation, as suggested in Ref. \cite{2006PhRvC..73a5801S}, remains
valid. 

It is important to note that the negative slop of the $S$ factor,
as determined by experimental methods, with respect to energy at small values of
$E$ is attributed to electron screening, as suggested in Ref. \cite{2021ApJ...920..145H}. At small energies, obtained partial ($^{7}$Li($d$,$n$)$^{8}$Be($0^{+}$)) and total ($^{7}$Li($d$,$n$)$\alpha\alpha$) $S$
factors fall between experimental data of Sabourov et al. and Hou et al., , steadily increasing with energy.%

\begin{figure}
\begin{center}
\includegraphics[width=\textwidth]{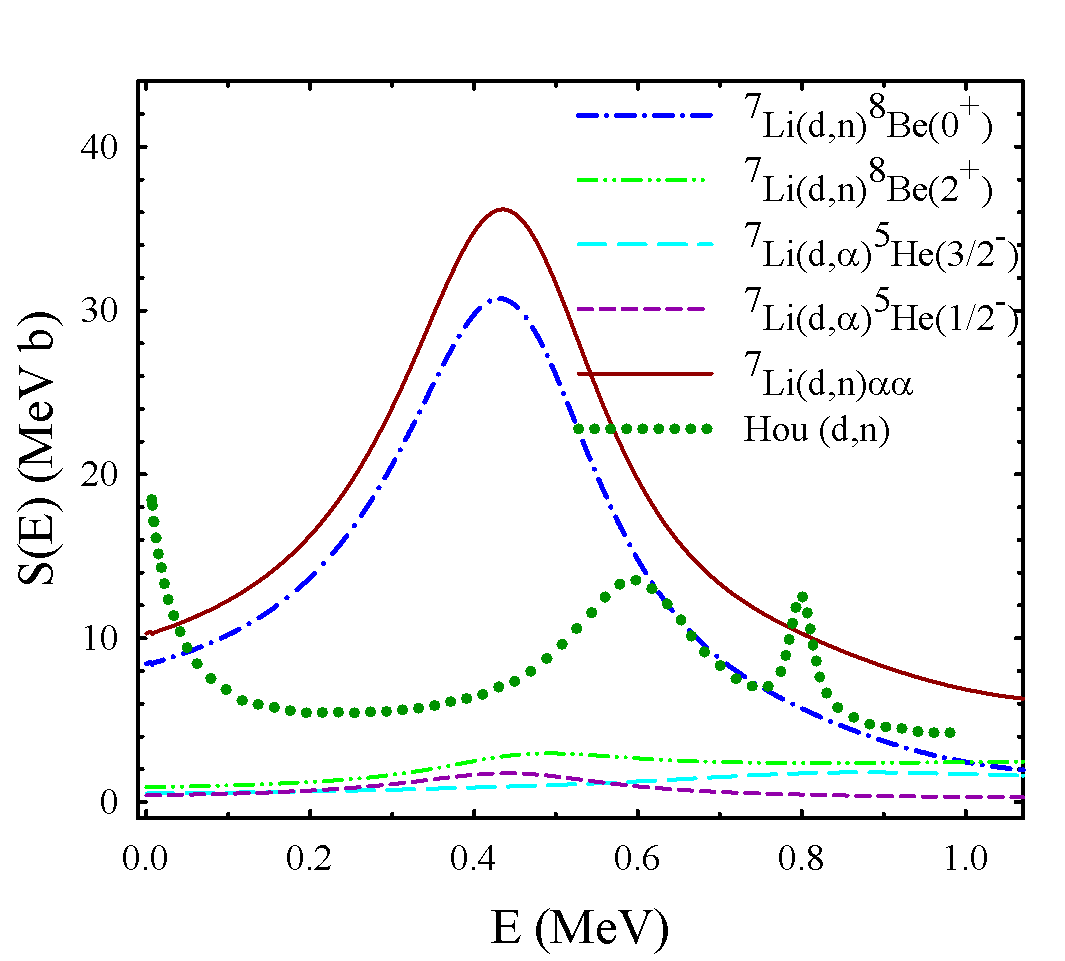}
\caption{Comparison of the experimental (Hou) and calculated astrophysical $S$
factor of the reaction generated by $^{7}$Li+$d$ collision}%
\label{Fig:Sfactors7LiDTvsE}%
\end{center}
\end{figure}

In Fig. \ref{Fig:Sfactors7LiDTvsE} we display astrophysical $S$ factors in
wider energy range and compare only with results of Hou et al. The
astrophysical $S$ factor of $^{7}$Li($d$,$n$)$^{8}$Be($0^{+}$) and $^{7}%
$Li($d$,$n$)$\alpha\alpha$, obtained in our model, have significant peak  at the
energy $E\approx$0.42 MeV, while the experimental peak with smaller amplitude is
observed at E$\approx$0.59 MeV. From an energy of approximately $E\approx$0.6 MeV onward, theoretical results closely align with experimental data, capturing the overall decreasing trend of $S$ factors, except for the second peak at $E\approx$0.8 MeV.

\section{Conclusions}
\label{concl}

We have applied a many-configurational microscopic model to study resonance
states in $^{9}$Be and $^{9}$B, focusing on the reactions $^{7}$Li$\left(  d,n\right)
\alpha\alpha$ and $^{7}$Be$\left(  d,p\right)  \alpha\alpha$. Two
three-cluster configurations $\alpha+\alpha+n$ and$\quad\alpha+d+^{3}$H in
$^{9}$Be, and $\alpha+\alpha+p$ and$\quad\alpha+d+^{3}$He in $^{9}$B, have been
used to invoke the dominant binary channels in $^{9}$Be and $^{9}$B, respectively.

Three-cluster configurations were projected on a set of four binary channels%
\begin{equation*}
^{8}\text{Be}+n\text{,}^{5}\text{He}+\alpha\text{,}^{7}\text{Li}+d\text{,
}^{6}\text{Li}+^{3}\text{H}%
\end{equation*}
in $^{9}$Be and four channels
\begin{equation*}
^{8}\text{Be}+p\text{,}^{5}\text{Li}+\alpha\text{,}^{7}\text{Be}+d\text{,
}^{6}\text{Li}+^{3}\text{He}%
\end{equation*}
in $^{9}$B. Besides, within the present model, clusters $^{5}$He, $^{5,6,7}$Li,
$^{7,8}$Be were considered as two-cluster systems.
Dominant two-cluster structures were employed for these nuclei. Clusters
$^{6}$Li,$^{7}$Li,$^{7}$Be were represented by their bound states, while
clusters $^{5}$He and $^{8}$Be, which are unbound, were represented
by pseudo-bound states.

All calculations were performed with the Minnesota potential. Parameters
of the potential were carefully chosen to ensure that the energies of the 3/2$^{-}$
bound state in $^{9}$Be and  the low-lying 3/2$^{-}$ resonance states in
$^{9}$B closely matched the experimental data. This approach allowed for a determination of cluster-cluster
interactions  in the most appropriate manner.

We examined resonance states near the two-cluster thresholds: $^{6}$Li$+^{3}$H and $^{7}$Li$+d$ in $^{9}$Be, and $^{6}$Li$+^{3}$He 
and $^{7}$Be$+d$ in $^{9}$B. Our calculations revealed a significant number of resonance states, predominantly with negative parity. For each resonance state, we identified its dominant decay channels.

With the provided input parameters, we achieved a satisfactory description of certain high-lying resonance states. However, it's important to note that there are challenges in accurately reproducing resonance states around the $^{7}$Li+$d$ and $^{7}$Be+$d$ decay channels. The parameters for part of these states currently deviate from experimental values. Further refinement of our approach may be needed to enhance the agreement in these specific decay channels.

The observed disagreement between theoretical and experimental data can be attributed, in part, to the simplification of exit channels in the present model and chosen nucleon-nucleon potential. 

While our model treats all exit channels of reactions induced by $^{7}$Li+$d$ and $^{7}$Be+$d$ collisions as two-body channels, they are inherently three-body channels due to the lack of bound states in nuclei $^{8}$Be, $^{5}$He, and $^{5}$Li. These nuclei exhibit wide resonance states, except for the 0$^{+}$ resonance state, with total widths such as 1.51 MeV for the 2$^{+}$ resonance state in $^{8}$Be and 0.65 and 1.23 MeV for the 3/2$^{-}$ resonance states in $^{5}$He and $^{5}$Li, respectively.

To address this, we approximated these resonances using pseudo-bound states. We acknowledge that a more accurate three-body treatment of decay channels, specifically $^{8}$Be+$p$, $^{8}$Be+$n$, $^{5}$He+$\alpha$, and $^{5}$Li+$\alpha$, could enhance our description of resonance states in $^{9}$Be and $^{9}$B. This improvement, in turn, would likely contribute to a more accurate astrophysical $S$ factor for the reactions $^{7}$Li+$d=\alpha+\alpha+n$ and $^{7}$Be+$d=\alpha+\alpha+p$. We anticipate that incorporating weak coupling between two-body channels ($^{7}$Li+$d$, $^{7}$Be+$d$) and three-body channels ($\alpha+\alpha+n$, $\alpha+\alpha+p$) could lead to the emergence of narrow resonance states. However, undertaking such investigations would require substantial modifications to existing microscopic models.

We also explored the reactions $^{7}$Li$\left(d,n\right)\alpha\alpha$ and
$^{7}$Be$\left(d,p\right)\alpha\alpha$ in a two-body approximation for the exit
channels. 

Astrophysical $S$ factors of four reactions
\begin{equation*}
^{7}\text{Li}\left(  d,n\right)  ^{8}\text{Be}\left(  0^{+}\right)  ,\quad
^{7}\text{Li}\left(  d,n\right)  ^{8}\text{Be}\left(  2^{+}\right)  ,\quad
^{7}\text{Li}\left(  d,\alpha\right)  ^{5}\text{He}\left(  3/2^{-}\right)
,\quad^{7}\text{Li}\left(  d,\alpha\right)  ^{5}\text{He}\left(
1/2^{-}\right)
\end{equation*}
in $^{9}$Be and four reactions
\begin{equation*}
^{7}\text{Be}\left(  d,p\right)  ^{8}\text{Be}\left(  0^{+}\right)  ,\quad
^{7}\text{Be}\left(  d,p\right)  ^{8}\text{Be}\left(  2^{+}\right)  ,\quad
^{7}\text{Be}\left(  d,\alpha\right)  ^{5}\text{Li}\left(  3/2^{-}\right)
,\quad^{7}\text{Be}\left(  d,\alpha\right)  ^{5}\text{Li}\left(
1/2^{-}\right)
\end{equation*}
in $^{9}$B have been thoroughly investigated.  It was shown that at low
energy range these reactions are mainly proceed through the states 1/2$^{-}$
and 3/2$^{-}$. These two states generate large peaks of the $S$ factors as a
function of energy $E$ in the incident channel. These peaks are  related to
rather wide 1/2$^{-}$ and 3/2$^{-}$ resonance states. For example, parameters
of the 1/2$^{-}$ resonance states are $E$=0.45 MeV and $\Gamma$=0.68 MeV in
$^{9}$Be and  $E$=0.79 MeV and $\Gamma$=1.10 MeV in $^{9}$B. 

The total astrophysical $S$ factor for the $^{7}$Be($d$,$p$)$^{8}$Be(0$^{+}$) reaction, as predicted by our present model, demonstrates a favorable agreement with experimental data \cite{1960NucPh..18..492K}, particularly around the main peak situated at $E$=0.68 MeV. Furthermore, our results for the $^{7}$Be($d$,$p$)$^{8}$Be(2$^{+}$) reaction exhibit close alignment with experimental findings \cite{2019PhRvL.122r2701R} within the energy range of 0.3$<E<$0.75 MeV and beyond $E>$1.5 MeV.

Moreover, it's noteworthy that the theoretical $S$ factor for the $^{7}$Be($d$,$p$)$^{8}$Be(0$^{+}$) reaction features a peak at an energy proximate to the experimental peak observed in \cite{2019PhRvL.122r2701R}. However, it's important to acknowledge that the height of the theoretical peak is somewhat smaller than that of the corresponding experimental peak.

The astrophysical $S$ factors for $^{7}$Li($d$,$n$)$^{8}$Be($0^{+}$) and $^{7}$Li($d$,$n$)$\alpha\alpha$, as predicted by our model, exhibit a peak at a slightly lower energy compared to the experimental data \cite{2021ApJ...920..145H}. However, starting from an energy of approximately $E\approx$0.6 MeV, our theoretical results closely mirror the observed experimental trends, capturing the overall decreasing behavior of the $S$ factors.

It's noteworthy that, at small energies, both the partial ($^{7}$Li($d$,$n$)$^{8}$Be($0^{+}$)) and total ($^{7}$Li($d$,$n$)$\alpha\alpha$) $S$ factors fall within the range defined by the experimental data from \cite{2021ApJ...920..145H} and \cite{2006PhRvC..73a5801S}.

It's important to highlight that our calculations do not indicate the presence of any narrow resonance states near the $d+^{7}$Be threshold, especially within an energy range that is currently beyond the reach of modern experimental methods.

\begin{acknowledgments}

This work received partial support from the Program of Fundamental Research of the Physics and Astronomy Department of the National Academy of Sciences of Ukraine (Project No. 0122U000889). We extend our gratitude to the Simons Foundation for their financial support. Additionally, Y.L. acknowledges the National Institute for Nuclear Physics, Italy, for providing a research grant to support Ukrainian scientists.

\end{acknowledgments}


\end{document}